\documentclass[10pt,twocolumn,twoside]{IEEEtran}
\usepackage{amsmath,amscd,amssymb,amsgen,amsfonts,amsbsy}
\usepackage{mathrsfs}
\usepackage[utf8x]{inputenc}
\usepackage{color}
\usepackage{textcomp}
\usepackage{float}
\usepackage{latexsym,graphicx}
\usepackage{mathtools}
\usepackage{breqn}
\usepackage{tikz}
\usepackage{breqn}
\usetikzlibrary{automata,arrows,positioning,calc}
\usepackage{url}
\usepackage{cite}
\usepackage{algorithmic}
\usepackage[ruled,lined]{algorithm2e}

\newcommand{\bm}[1]{\boldsymbol{#1}}
\newcommand{\prob}[1]{\mathsf{Pr}\left( #1 \right)}

\newcommand{\remove}[1]{}

\newcommand{\comments}[1]{}

\newcommand{\qed}{\hfill $\square$}

\newtheorem{corollary}{Corollary}
\newtheorem{lemma}{Lemma}

\newtheorem{theorem}{Theorem}

\newtheorem{remark}{Remark}

\newtheorem{definition}{Definition}
\newtheorem{result}{Result}
\makeatother  

\title{ Constrained Restless Bandits for Dynamic Scheduling in Cyber-Physical Systems }

\author{
 $\text{Kesav Kaza}^{}$, Rahul Meshram, Varun Mehta and S.~N.~Merchant \\
  \thanks{ The major part of this work was done when all the authors were with the Department of Electrical Engineering, Indian Institute of Technology Bombay, Mumbai, India. Kesav Kaza is now with Polytechnique Montreal, Canada. Rahul Meshram is now with IIIT Allahabad, India. Varun Mehta is now with University of Ottawa, Canada. S. N. Merchant is with IIT Bombay, India.}}

\begin{document}
%

\maketitle

\begin{abstract}
This paper studies a class of constrained restless multi-armed bandits (CRMAB). The constraints are in the form of time varying set of actions (set of available arms). This variation can be either stochastic or semi-deterministic. Given a set of arms, a fixed number of them can be chosen to be played in each decision interval. The play of each arm yields a state dependent reward.
The current states of arms are partially observable through binary feedback signals from arms that are played. The current availability of arms is fully observable.  
 The objective is to maximize long term cumulative reward. 
The uncertainty about future availability of arms along with partial state information makes this objective challenging. Applications for CRMAB can be found in resource allocation in  cyber-physical systems involving components with time varying availability.
 
First, this optimization problem is analyzed using Whittle's index policy. To this end, a constrained restless single-armed bandit is studied. It is shown to admit a threshold-type optimal policy and is also \textit{indexable}. An algorithm to compute Whittle's index is presented. An alternate solution method with lower complexity is also presented in the form of an online rollout policy. A detailed discussion on the complexity of both these schemes is also presented, which suggests that online rollout policy with short look ahead is simpler to implement than Whittle's index computation. Further, upper bounds on the value function are derived in order to estimate the degree of sub-optimality of various solutions. 
The simulation study compares the performance of Whittle's index, online rollout, myopic and modified Whittle's index policies. 

\end{abstract}


\IEEEpeerreviewmaketitle

\section{Introduction}
Restless multi-armed bandits (RMABs) are a class of discrete-time stochastic control problems which involve sequential decision making with a finite set of actions (called arms). RMABs are used in applications involving decision making under uncertainty in evolving environments. 
They have been extensively studied for scheduling
applications in opportunistic communication systems, dynamic relay selection in  wireless relay networks, queuing systems, multi-agent systems, recommendation systems, unmanned aerial vehicle routing, internet of things, scheduling machine maintenance and cyber-physical systems \cite{LiuZhao10, Kaza19, Kaza2018wcnc, Gittins11, JWang19, Ansell03,Nino-Mora01,Nino-Mora07, Avrachenkov18, Meshram17,Le08,akbarzadeh2021maintenance,Guo18}.

\begin{figure}
	\begin{center}
		\begin{tabular}{lr}
			  \label{fig:cps-flow}
			\includegraphics[width=0.4\columnwidth]{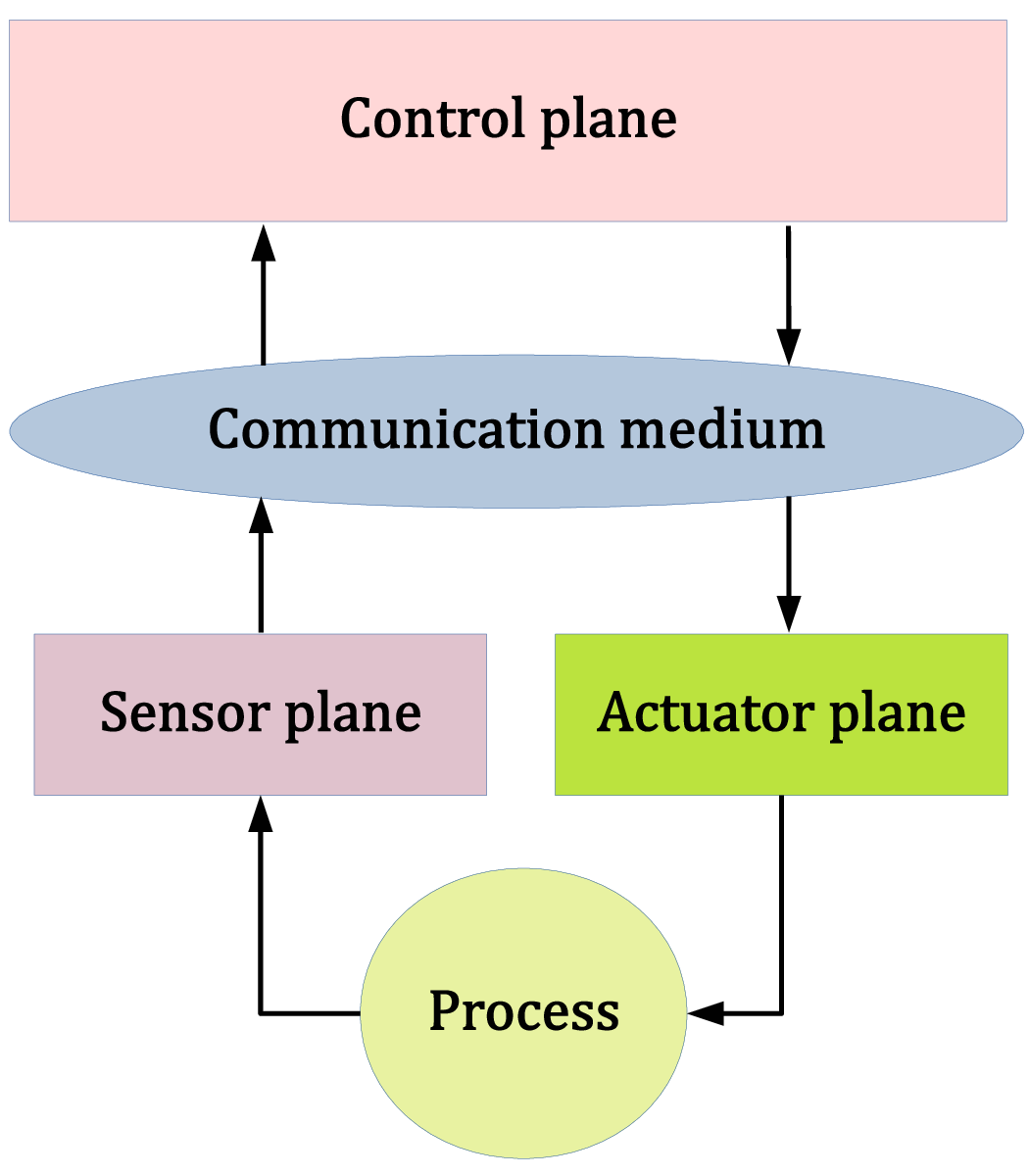}
			&   
			\includegraphics[width=0.54\columnwidth]{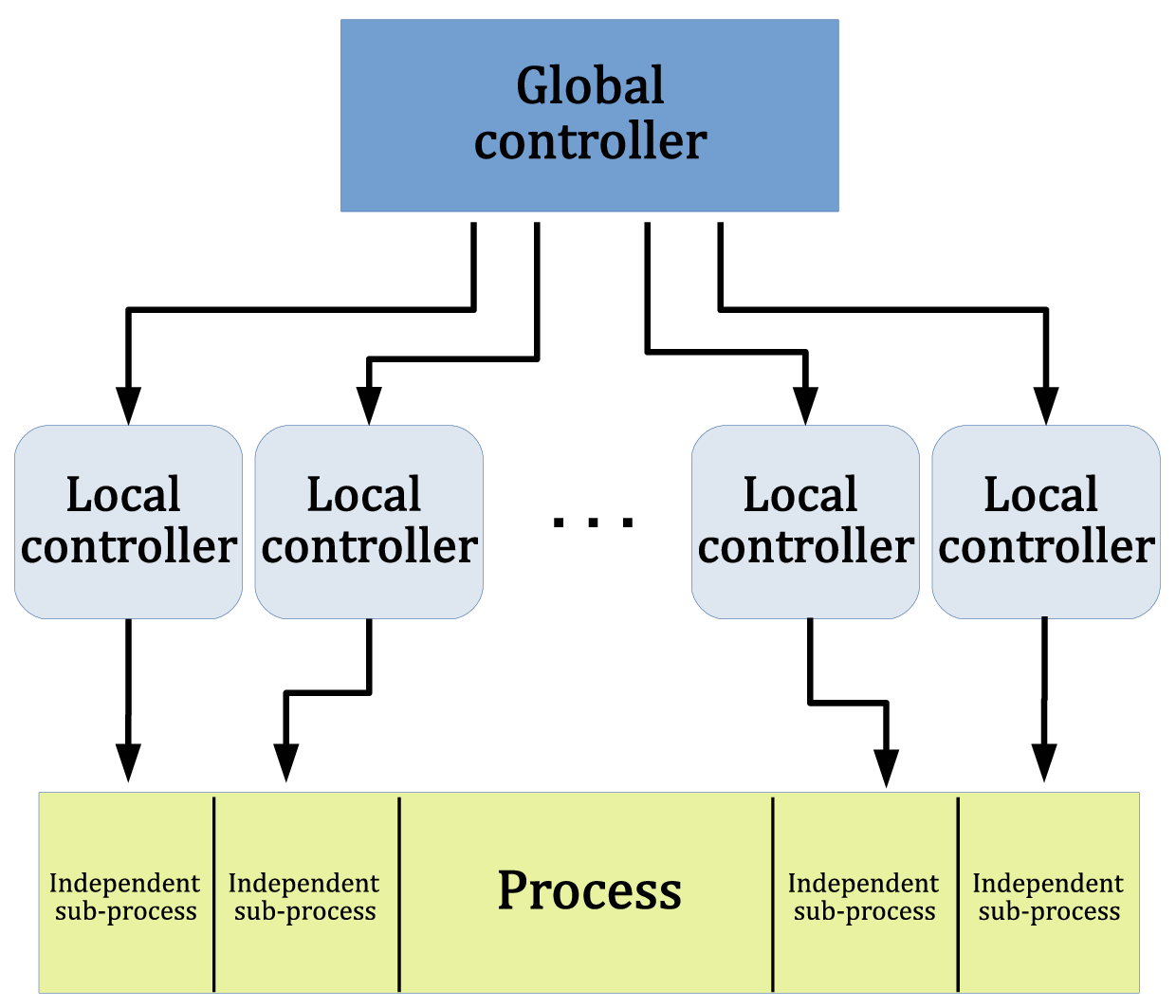} \\
		\end{tabular}
	\end{center}
	\caption{A simplified schematic showing the major parts of a cyber-physical system. \emph{(left)} Information flow in a CPS. \emph{(right)} A layered CPS used to control a complex process with independent sub-processes.}
	\label{fig:cps-layered}
\end{figure}

Let us first look at some application scenarios involving dynamic scheduling in uncertain environments under resource constraints.
Cyber-physical systems (CPS) have received much attention in the recent times in view of their potential applications relating to environment, health care, security, etc \cite{Kim13overview,Guan16}. 
A simplistic representation of the interaction between various elements of a CPS is shown in Fig.~\ref{fig:cps-layered}. 
%
%
%
In highly complex cyber-physical systems, there may be layers of decision making involving local controllers. 
Here, the global controller's goal is to control a complex process with many independent sub-processes, for maximizing its rewards. For this purpose a local controller is assigned to interact with each sub-process. The global controller aims for macro-optimization while the local controllers are tasked with micro-optimization. 
%
%

{Clearly, in these systems the problem of scheduling and resource allocation occurs at multiple places such as sensor scheduling for monitoring of dynamic processes, task scheduling for control of sub-processes, etc.
%
%
%
Further, these scheduling problems often come with resource availability and latency constraints. 
For example, some of the local controllers may not be available in certain time slots on account of their being engaged or in maintenance. This availability may be time varying and depend on factors such as maintenance time. Similarly, communication channels might become unavailable in some time slots due to heavy interference. 

Let us consider the problem of sensor scheduling in CPS. There are $N$ dynamic processes which are to be monitored using one sensor each. The sensors need  to transmit data to a controller over a wireless network with limited number of channels $(<N)$ which are evolving and hidden from the controller.  In such cases, the controller only has partial knowledge of the system state. For effective monitoring of the system, sensors need to be scheduled appropriately by the controller such that a long term objective function can be maximized. The scheduling scheme must also consider the fact that sensors have energy constraints and may intermittently be unavailable for transmission.} 
In this paper, we formulate the problem of scheduling under dynamic resource constraints using restless multi-armed bandit (RMAB) models.
%
%
%
%
In the following discussion we provide a brief overview of RMABs.
\subsection{Restless multi-armed bandits}
An RMAB is described as follows. There is a decision maker or source that has $N$
independent arms. Each arm can be in one of a finite set of states and the
state evolves according to Markovian law. The play of an arm yields a
state dependent reward. It is assumed that the decision maker knows
the statistical characteristics of state evolution for each arm. The
system is time-slotted. The decision maker
plays $M$ out of $N$ arms in each slot. The goal is to determine
the sequence of plays of the arms that maximizes the long term
cumulative reward. These planning problems are non-trivial because
there is a trade-off between the immediate and future rewards. The
choice that yields low immediate reward may yield better future reward.

A typical RMAB model assumes that the arms are
always available and the objective is to determine the optimal subset of arms to be played in a given state. We consider the case where the
availability of arms is intermittent and time varying. We refer to such RMABs as \textit{constrained restless multi-armed bandits (CRMAB)}. This is a generalization of restless multi-armed bandits. Models for availability of arms may vary across applications. We consider stochastic and semi-deterministic availability models.
 
{Apart from scheduling in CPS, another potential application for CRMABs is the problem of dynamic relay selection in wireless networks with evolving channel conditions and intermittent availability of relays due to energy constraints or other application interrupts.}
\subsection{Related work}
The literature on restless bandits is vast and includes different
variations on bandits and their applications. We mention 
a few that are relevant to our work.

The restless multi-armed bandit problem was first proposed in 
\cite{Whittle88}. It was inspired from the work on rested bandits \cite{Gittins79}.  In \cite{Gittins79}, index policies were introduced for rested
multi-armed bandits, where states of arms are frozen when they
are not played. This index policy is now known as Gittins index
policy. Later, \cite{Whittle88} studied restless bandits and
introduced a heuristic index policy which is now referred to as Whittle's index
policy. The popularity of Whittle's index policy is due to its asymptotic optimality in some examples and its near
optimal performance in some others (see \cite{Whittle88, Weber90, Ouyang12}).
 Whittle's index policy for other applications such as machine maintenance and repair problems are analysed in \cite{Gittins11,akbarzadeh2021maintenance}. Recently, \cite{Guo18} applied the RMAB framework to the problem of risk-sensitive scheduling in CPS. Here, an exponential cost function is defined instead of a linear function. This variant is termed as risk-sensitive RMAB, and the corresponding index policy as risk-sensitive index policy. 


In classical restless bandit literature, current states of all the arms are observable in every time slot \cite{Nino-Mora01, Nino-Mora07, Whittle88}.  Later, this assumption was relaxed and restless bandit models with partially observable states were studied, where states are observable only for those arms that are played \cite{LiuZhao10,Ouyang12}. Recent work on restless bandits further generalized this model to the case where states of all arms are partially observable. This is referred to as the \textit{hidden restless bandit} \cite{Borkar17,Meshram18}. In \cite{Kaza19}, further generalization is considered where multiple state transitions are allowed in a single decision interval. More recently, \cite{akbarzadeh2021maintenance} considers the problem of multi-state RMAB and presents optimal threshold policy and indexability results under some structural assumptions.

Whittle's index policy for RMABs was studied for job scheduling and dynamic
routing on servers in \cite{Martin05, Glazebrook07}, where authors
considered scenario of servers being available intermittently.

{An RMAB variant with availability constraints on arms was first proposed in \cite{Dayanik02}. It was applied to the machine repair problem  where machine availability is time varying and machine state is observable. This paper introduced two models---1) Unavailable arms are cannot played, 2) Unavailable arms can be played with some additional penalty.  Whittle's index policy is studied.  Since states are exactly observable, a closed form expression for the index is easily obtained.   
The second model of \cite{Dayanik02} was generalized for partially observable states  by \cite{Varun18}. In this model too there is a penalty for playing an unavailable arm, i.e., arms can be played both when they are available and unavailable. Whittle's index policy and myopic policy are analyzed.
 CRMABs considered in the current paper do not allow to play unavailable arms and exact state of arms is not observable. Further, we also consider a semi-deterministic availability  model. The results claimed in \cite{Varun18} do not necessarily apply to the current model because of difference in belief update rules and assumptions on reward structure. }

%


The literature on POMDPs, RMABs makes use of certain common techniques and procedures. These include defining action value functions and using induction principle to derive their structural properties. Another common aspect is proving sub-modularity of the value function, which will lead to a threshold structure of optimal policy (see \cite{Lovejoy87, Puterman14}). The threshold structure of the optimal policy is used to prove indexability of restless single armed bandits. This allows one to apply  Whittle's index policy. One must note the differences in modeling that require redoing or following similar procedures as it is not obvious that the same results hold.

\subsection{Contributions}
We consider the problem of restless multi-armed bandit with dynamic resource constraints and partially observable states. It is referred to as partially observable CRMAB. We study two availability  models --- stochastic and semi-deterministic.

We analyze the constrained restless single-armed bandit (CRSAB) problem, show the threshold sturcture of  optimal policy and thereby indexability. We present an algorithm for Whittle's index computation which is based on two-timescale stochastic approximation (TTSA). We also present an online rollout policy with a simpler implementation, as an alternative to Whittle's index policy.

A detailed discussion on sample and computational complexity of Whittle's index and online rollout policy is presented. Sample complexity results (time to convergence to $\epsilon-$optimal answer) for TTSA remains an open problem. We present this result (conjecture) by drawing parallels between TTSA and the primal-dual schemes used to solve constrained MDP problems (see Section~\ref{sec:sample_complexity}).

{ 
As the optimal solution to the CRMAB problem is difficult to obtain, and both Whittle's index policy and online rollout policy are heuristic approaches, we derive an upper bound on the optimal value function of the original problem. In this, we use the idea of Lagrangian relaxation of the problem from \cite{Adelman08}. This bound can be used to measure the degree of sub-optimality of various policies. We further study the relationship between Lagrangian bound for CRMAB and unconstrained RMAB. It is shown that under certain conditions the former gives a tighter bound than the latter.
}


Finally, a simulation study is presented with performance comparison of various policies. We observe that online rollout policy sometimes performs better than Whittle's index policy. Further, both these policies are better than the myopic policy.


The rest of this document is organized as follows. The system model is explained in Section~\ref{sec:model} and the constrained restless single armed bandit is analyzed in Section~\ref{sec:crsab}. Sample complexity results are discussed in Section~\ref{sec:sample_complexity}. Online rollout policy is presented in Section~\ref{sec:rollout}. Bounds on value functions are derived in Section~\ref{sec:bounds}. Numerical simulations are presented in Section~\ref{sec:simulation}, and concluding remarks in Section~\ref{sec:conc}.

\section{Model Description and Preliminaries}
\label{sec:model}
Consider a restless multi-armed bandit with $N$ independent arms.
Each arm can be one of two states, state $0$ or state $1$. 
The state of each arm evolves according to a discrete time Markov chain.
Some times the arms might become unavailable. 
The evolution of states also depends on the availability of arms. The availability of arms is time varying. 
 Let us introduce some notation to formalize the model.
The system is time-slotted and time is indexed by
$t.$ Let $X_n(t) \in \{0,1\}$ denote the state of arm $n$ at the beginning of time slot $t.$ 
 
Let $Y_n(t) \in \{0,1\}$ denote the availability of arm $n$ at the
beginning of time slot $t$ and
%
\begin{eqnarray*}
 Y_n(t) = 
 \begin{cases}
 1 & \mbox{if arm $n$  is available,} \\
 0 & \mbox{if arm $n$ is not available.}
 \end{cases}
\end{eqnarray*}  
%
Each arm has two actions associated with
it when it is available, either `play' or `don't play'. When it is unavailable it cannot be played. However, it's state still evolves.
Let $a_n^{1}(t) \in \{0,1 \}$ denote the action corresponding to arm
$n$ when it is available, it is described as follows.
 %
\begin{eqnarray*}
  a_n^{1}(t) = \begin{cases}
    1 & \mbox{if arm $n$ is available and played,}\\
    0 & \mbox{if arm $n$ is available and not played.} 
  \end{cases}
\end{eqnarray*} 
%
Let $a_n^{0}(t)$ be the action corresponding to arm $n$ when it is not
available.  As it cannot be played, $a_n^{0}(t) \coloneqq 0.$ 
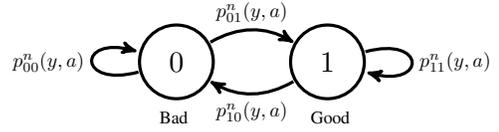
\begin{figure}
\centering
  \begin{center}
  \resizebox{0.37\textwidth}{!}{
    \begin{tikzpicture}[draw=black,>=stealth', auto, ultra thick, node distance=1.8cm]
      \tikzstyle{every state}=[fill=white,draw=black,ultra
        thick,text=black,scale=1.5] \node[state] (A) {$0$};
      \node[state] (B)[ right of=A] {$1$}; \path (A) edge[loop left]
      node{$p^n_{00}(y,a)$} (A) edge[bend left,above,->]
      node{$p^n_{01}(y,a)$ } (B) (B) edge[loop right]
      node{$p^n_{11}(y,a)$} (B) edge[bend left,below,->]
      node{$p^n_{10}(y,a) $} (A); 
      \draw (-0.1,-1.0) node{\small{ Bad}};
      \draw (2.7,-1.0) node {\small{ Good }};
    \end{tikzpicture}
		}
    \caption{ Two state channel model along with the
      notation of transition probabilities for $n^{th}$ arm.}
\label{fig:Arm-MC}   
\end{center}  
\end{figure}

The state of arm $n$ changes at beginning of time slot $(t+1)$ from
state $i$ to $j$ according to transition probabilities
$p_{ij}^n.$  These are defined as follows.
%
\begin{eqnarray*}
p_{ij}^n \coloneqq \Pr \{ {X_n}(t + 1) = j~|~{X_n}(t) = i\}.
\end{eqnarray*}
%
When arm $n$ is played, the result is either success or
failure. A binary signal is observed at the end of each slot that
describes the event of success or failure (ACK or NACK in communication parlance). Let $Z_n(t)$ be the binary signal that is received by the source (decision maker) at the end of slot $t.$ It is given as %
 \begin{eqnarray*}
 Z_n(t) = 
 \begin{cases}
 1 & \mbox{if arm $n$ is played and that resulted success,} \\
 0 & \mbox{If arm $n$ is played and no success.}
 \end{cases}
 \end{eqnarray*}
 %
When arm $n$ is not played, no signal is observed from that arm. 
Let  $\rho_n(i)$ be the probability of success from playing arm $n.$\\
 $\rho_n(i) \equiv \rho_{n,i} := \Pr{\left(Z_n(t) = 1~|~X_n(t) = i, a_n(t) =1 \right)},
$
for $i = 0,1.$ It is the probability that signal $Z_n(t)=1$ is observed given that arm $n$ is in state $i$ and action $a_n(t) =1.$ We will assume $ \rho_{n,0} <
\rho_{n,1},$ i.e., the probability of success is higher from state
$1$ than from state $0.$
 
The play of arm $n$ yields a state dependent reward. Let $\eta_{n,i}$
be the reward obtained by playing arm $n$ given that $X_n(t) = i.$
 When arm $n$ is not played, no reward is obtained.
Further, we suppose that $0 \leq \eta_{n,0} < \eta_{n,1} \leq 1$ for
all $n.$

The decision maker or source cannot observe the exact state vector at any arbitrary time $t$. 
However, it can exactly observe the current
availability vector at the beginning of each time slot. That is,
 $\bm{Y}(t) = [Y_1(t),...,Y_n(t)]$ is known at beginning of slot $t.$ 
Since the source does not know the exact states of arms, it maintains
a `belief' about each of them. Let $\pi_n(t)$ be the belief about arm
$n.$ It is the probability of being in state $0,$ given the history
$H_t$ upto time $t.$ \\
The history upto time $t$ is given as 
\begin{align*}
H_t := \left(Y_n(s),a_{n}(s) ,Z_{n}(s) \right)_{0 \leq n \leq N, 1 \leq s < t}.
\end{align*}
The belief vector is given as $\boldsymbol{\pi}(t) =[\pi_1(t),...,\pi_n(t)],$ with
$\pi_n(t) = \Pr{\left( X_n(t) = 0~|~ H_t  \right)}.
$
\subsection{Availability models}
We consider two availability models, namely, stochastic and semi-deterministic. In the stochastic model, future availability depends on a probability value conditioned on current availability.
 In the semi-deterministic model, future availability is deterministic when an arm goes unavailable. This model is useful in applications in which some sub-systems are occasionally down for a fixed period of time for maintenance.
\subsubsection{Stochastic}
The future availability of arm $n,$  $Y_n(t+1),$ is dependent on current availability $Y_n(t)=y,$ action $a_n(t)=a$ and current state of arm ${X_n}(t) = i.$ We define
{{
\begin{equation*}
	\theta_n^a(i,y) := \Pr{ \left({Y_n}(t+1) = 1|{X_n}(t) = i ,{Y_n}(t) =
		y,a_n(t) = a \right) }.
\end{equation*}
}}
We replace knowledge of state ${X_n}(t) = i$ with belief $\pi_n(t) = \pi,$ and we rewrite $ \theta_n^a(i,y) $ as $\theta_n^a(\pi,y).$ 
The availability model is described as follows.
\begin{equation*}
 Y_n(t+1) = 
 \begin{cases}
 & 1, \texttt{ } w.p. \texttt{ } \theta_n^a(\pi,y),\\
  & 0, \texttt{ } w.p. \texttt{ } 1-\theta_n^a(\pi,y).
 \end{cases}
\end{equation*}
The decision maker knows the probability of availability $\theta_n^a(\pi,y).$ Notice that this model satisfies Markov property.

In general, $\theta_n^a(i,y)$ depends on the state of arm
$n,$ current availability $y$ and action of that arm $a.$ 

{The model for $\theta_n^a(\pi,y)$ might depend on application. For simplicity, we assume that $\theta_n^a(\pi,y)$ to be linearly dependent on $\pi$ and $y$ for each $a \in \{0,1\}.$}

\subsubsection{Semi-deterministic} 
The future availability for unavailable arms has a deterministic model. When available arms turn unavailable, they remain unavailable for exactly $T_0$ slots and then   become available. 
That is,\\
if arm $n$ is unavailable, i.e.,  $Y_n(t)=0,$ then 
\begin{eqnarray}
 Y_n(t+t') = \begin{cases}
             & 0, \texttt{ } for \texttt{ } t' = 1,...,T_0-1, \nonumber\\ 
	     	 & 1, \texttt{ } for \texttt{ } t' = T_0.
            \end{cases}
\end{eqnarray}
If arm $n$ is available, i.e.,  $Y_n(t)=1,$ then
\begin{equation*}
 Y_n(t+1) = 
 \begin{cases}
 & 1, \texttt{ } w.p. \texttt{ } \theta_n^a(\pi,1),\\
  & 0, \texttt{ } w.p. \texttt{ } 1-\theta_n^a(\pi,1).
 \end{cases}
\end{equation*}
%
\subsection{Problem formulation}
As the exact state is not observable the state is redefined in terms of belief and availability. Consider the perceived state $S_n(t) = (\pi_n(t), Y_n(t)) \in [0,1]
\times \{0,1\}$ in beginning of time slot $t.$ Using the belief
$\pi_n(t),$ we compute the expected reward from play of arm $n$ at
time $t$ as follows. 
\begin{equation*}
\eta(\pi_n(t), y=1) \coloneqq \pi_n(t) \eta_{n,0} + (1-\pi_n(t)) \eta_{n,1}
\end{equation*}
and $\eta(\pi_n(t), y=0)\coloneqq 0.$ 

We next define the optimization problem as reward maximization.  Let
$\phi(t)$ be the policy of the source such that $\phi(t): H_t
\rightarrow \{1, \cdots,N\}$ maps the history to $M$ arms in slot $t.$
Let
\begin{eqnarray*}
a_n^{\phi}(t) = 
\begin{cases}
1 & \mbox{if $n \in \phi(t),$ } \\
0 & \mbox{if $n \notin \phi(t).$}
\end{cases}
\end{eqnarray*} 
The infinite horizon discounted cumulative reward under strategy
$\phi$ for initial state information $(\bm{\pi},
\bm{y}),$ $\bm{\pi}=(\pi_1(1), \cdots, \pi_N(1))$ and
$\bm{y} = (y_1(1), \cdots, y_N(1))$ is given by
\begin{dmath}
V_{\phi}(\bm{\pi}, \bm{y}) = 
\mathrm{E}^{\phi}\left({\sum_{t=1}^{\infty} \beta^{t-1} 
\left[ \sum_{n=1}^{N} 
a_n^{\phi}(t) \eta(\pi_n(t), Y_n(t) ),
\right]
}\right),\\
{\texttt{\hspace{1 cm}}\sum_{n=1}^{N} a_n^{\phi}(t) = M.}\nonumber
\label{eqn:opt1}
\end{dmath}  
In each time slot $M$ arms are played; hence the constraint $\sum_{n=1}^{N}
a_n^{\phi}(t) = M.$ Here, $\beta \in (0,1)$ is the discount
parameter. The objective is to find a policy $\phi$ that maximizes
$V_{\phi}(\bm{\pi}, \bm{y})$ for all $ \bm{\pi}
\in [0,1]^N,$ $\bm{y} \in \{0,1\}^N.$ The problem
\eqref{eqn:opt1} is a constrained hidden Markov restless multi-armed
bandit. 
{The optimal solution for problem \eqref{eqn:opt1} is computationally intractable; it is known to be PSPACE-hard  \cite{Papadimitriou99}. The major difficulty here is due to  the integer constraint, $\sum_{n=1}^{N}a_n^{\phi}(t) = M,$ $a_n^{\phi}(t) \in \{0,1\}.$  The key idea is to introduce a relaxed version of problem \eqref{eqn:opt1}. This is done by replacing the exact integer constraint with the following expectation constraint.} 
\begin{eqnarray}
\mathrm{E}^{\phi} \left(\sum_{t=0}^{\infty}\beta^{t-1} \left[ \sum_{n=1}^{N}
a_n^{\phi}(t)  \right] \right)= \frac{M}{1-\beta}.
\label{eqn:opt2}
\end{eqnarray} 

Now, using Lagrangian relaxation of the problem, 
we can reduce the dimension of the relaxed RMAB problem into $N$ restless single-armed  bandits. In the next section we study the constrained restless single-armed  bandit problem and define an index policy.

{Note: The above formulation assumes that at least $M$ are available in each slot. When the number of available arms (say $m$) in a slot is less than $M,$ then $M-m$ dummy arms with minuscule rewards (say $\epsilon_i>0$ for state $i$) are played.}

\section{Constrained restless single armed bandit}
\label{sec:crsab}

As there is only one arm, the problem of the decision maker here is to decide in each time slot whether or not to play the arm. We drop the subscript $n$, the sequence number of the
  arms; so, $\rho_{n,i} \equiv \rho_i, $ $\eta_{n,i} \equiv \eta_i,$ $\theta_n^{a}(y) \equiv \theta^{a}_{y}.$   
  The analysis of the single arm problem proceeds by  assigning a subsidy $w$ for not playing the arm.

Recall that the source maintains and updates its belief about state of the arm at the end of every time slot. The update rules are based on previous actions, availability and observations of the arm and it is given as follows.
\begin{enumerate}
 \item If the arm is available, played and a success is observed, i.e., 
 $a(t)=1, Y(t)=1, Z(t)=1.$ Then the new belief $\pi(t+1) =
 \Gamma_{1}(\pi(t)),$ and it is   
\begin{equation*}
 \Gamma_{1}(\pi) = \frac{\pi \rho_0 p_{00} + (1-\pi) \rho_1
   p_{10}}{\pi \rho_0 + (1-\pi)\rho_1}.
\end{equation*}
This update is according to the Bayes rule.
\item If the arm is available, played and success is not observed, i.e.,  $a(t)=1, Y(t)=1, Z(t)=0,$ then  the  belief $\pi(t+1) =
  \Gamma_{0}(\pi(t)),$ and it is 
\begin{equation*}
 \Gamma_{0}(\pi) = \frac{ \pi (1-\rho_0)p_{00} +
   (1-\pi)(1-\rho_1)p_{10} }{ \pi(1-\rho_0) + (1-\pi)(1-\rho_1) }.
\end{equation*}
\item If the arm is available  but not played, there is no observation, i.e., 
 $a(t)=0, Y(t)=1.$ Then the belief $\pi(t+1) =
  \gamma^{0}_{1}(\pi(t))$ and it is given by 
\begin{equation*}
 \gamma^0_1(\pi) = \pi p_{00} + (1-\pi)p_{10}.
\end{equation*}
\item If the arm is not available, then it can  not be played and no observation is available, i.e.,  $a(t)=0, Y(t)=0.$  We consider the belief $\pi(t+1) =
  \gamma^{0}_{0}(\pi(t))$ and it is updated according to following rule.
\begin{equation*}
 \gamma^0_0(\pi) = \frac{p_{10}}{p_{01}+p_{10}}  \ \ \ \ \text{or} \ \ \ \  \pi p_{00} + (1-\pi)p_{10}.
 \end{equation*}
In this case, belief is either taken to be the stationary probability or the value obtained by natural evolution of the Markov chain.
\end{enumerate}
 
\subsection{Value functions} 
 Given an state $(\pi,y),$ let $V(\pi,y)$ denote the expected cumulative discounted reward achieved by the optimal policy. $V$ is called the optimal value function.  
Let us now define the values of different actions depending on the belief and availability, in terms of $V$. The value for action $a,$ given belief $\pi$ and availability $y$ is denoted as $\mathcal{L}^aV(\pi,y)$ for $a\in\mathcal{A}_y,$ $y \in \{0,1\}.$ Here, $\mathcal{L}^aV$ is called action value function. $\mathcal{A}_y$ is the set of possible actions for availability $y.$ For our model, we have $\mathcal{A}_{1} = \{0,1\}$ and $\mathcal{A}_{0} = \{0\}.$ When the arm is unavailable ($y=0$), it cannot be played. 

The value functions for stochastic availability model are given as follows. 
\begin{itemize}
 \item[a)]  For action $a=1,$ and availability $y=1:$
{{
\begin{dmath*}
\mathcal{L}^{1}V(\pi,1) = \eta(\pi) +  \beta\rho(\pi)\left[ \theta^{1}_{1}(\pi)V(\Gamma_{1}(\pi),1)+ (1-\theta^1_1(\pi))V(\Gamma_{1}(\pi),0) \right] + 
 \beta(1-\rho(\pi))\left[ \theta^1_{1}(\pi)V(\Gamma_{0}(\pi),1) + 
  (1-\theta^1_{1}(\pi))V(\Gamma_{0}(\pi),0) \right]
\end{dmath*}
 }}
Here, $\eta(\pi) = \eta_0 \pi + \eta_1 (1-\pi).$ 
 
The value function consists of immediate expected reward and discounted future value. So, the first term is immediate reward, $\eta(\pi).$ The second term and third terms depend on probability of observing success or failure. These terms also include the future value function and expectation w.r.t. availability probability. 
\item [b)] For action $a=0,$ and availability $y=1:$
{{
\begin{dmath*}
\mathcal{L}^0V(\pi,1) = w + \beta
\left[ \theta^0_1(\pi) V(\gamma^0_1(\pi),1) +
  (1-\theta^0_1(\pi))V(\gamma^0_1(\pi),0) \right]
\end{dmath*}
}}
If the arm is available and is not played, the immediate reward is a subsidy $w.$  The second term of the value function includes the expectation of future value  which depends on availability probability and updated belief. 
\item[c)] Action $a=0,$ availability $y=0,$
{{
\begin{dmath*}
\mathcal{L}^0V(\pi,0) = w + \beta\left[ \theta^0_{0}(\pi) V(\gamma^0_0(\pi),1) +
   (1-\theta^0_0(\pi))V(\gamma^0_0(\pi),0) \right]       
\end{dmath*}
}}
This value function is very similar to preceding case. If the arm is unavailable, it cannot be played. The value function consists of immediate reward as subsidy $w$ and the expected future value which depends on availability probability and updated belief. This updated belief could stationary probability or the value obtained by natural evolution of the Markov chain. 
\end{itemize}
 
We will now write down action value function expressions for the semi-deterministic availability model. Recall that when arm is not available then it cannot be played for a fixed amount of time $T_0.$ Thus, the value function differs from earlier stochastic model for availability $y=0$ and action $a=0.$ The value function for availability $y =1$ and action $a=0$ or $a=1$ is similar to that of stochastic availability model. The value functions are given as follows.
%
\begin{itemize}
 \item[a)] Action $a=1,$ and availability $y=1:$
{{
\begin{dmath*}
\mathcal{ L}^{1}V(\pi,1) = \eta(\pi) +  \beta
\rho(\pi)\left[ \theta^{1}_{1}(\pi)V(\Gamma_{1}(\pi),1)+ (1-\theta^1_1(\pi))V(\Gamma_{1}(\pi),0) \right] \\ + 
 \beta(1-\rho(\pi))\left[ \theta^1_{1}(\pi)V(\Gamma_{0}(\pi),1) + 
  (1-\theta^1_{1}(\pi))V(\Gamma_{0}(\pi),0) \right] 
\end{dmath*}
}} 
\item[b)] Action $a=0,$ and availability $y=1:$
{{
\begin{dmath*}
\mathcal{L}^0V(\pi,1) = w + \beta
\left[ \theta^0_1(\pi) V(\gamma^0_1(\pi),1) +
  (1-\theta^0_1(\pi))V(\gamma^0_1(\pi),0) \right]
\end{dmath*}
}}
\item[c)] Action $a=0,$ and availability $y=0:$
 \begin{dmath*}
\mathcal{ L}^{0}V(\pi ,0) = w\frac{(1-\beta^{T_0})}{(1-\beta)} + \beta^{T_0}V\left( (\gamma^0_0)^{T_0}(\pi),1 \right)
\end{dmath*}

Since the arm is unavailable for $T_0$ number of slots, the discounted reward obtained in this period is $w \frac{(1-\beta^{T_0})}{(1-\beta)}.$ The second term is future discounted value after $T_0$ slots when the arm becomes available.

\end{itemize} 

Observe that there is no available choice of actions for $y=0.$ However, the value function $\mathcal{L}^0V(\pi,0)$ is important as it impacts other value functions.

The optimal value function $V$ satisfies the following dynamic programming optimality equations. 
\begin{dmath}
V(\pi,y) = \max_{a\in \mathcal{A}_y} \mathcal{L}^{a}V(\pi ,y),\\ { \forall \pi \in [0,1] \ \text{  and } \  y \in \{0,1\} .}
\label{eqn: Valfunc_def}
\end{dmath}
%

Note that we sometimes use the notation $\mathcal{L}_w^aV(\pi,y)$ in place of $\mathcal{L}^aV(\pi,y)$ to emphasize the dependence on $w.$ 
We obtain all results assuming $\eta_0 = \rho_0,$ and $\eta_1 = \rho_1.$

%
%

\subsection{Structural results}
 
 In the following, we derive structural results for value functions in the case of  stochastic availability. These results also hold true for the semi-deterministic availability model because the value functions are similar except at availability $y=0.$ We first define a threshold type policy.

\begin{definition}(Threshold type policy)
 A policy is said to be of threshold type if one of the following is true.
 \begin{enumerate}
 \item For $y =1$ and $\forall  \pi \in [0,1],$ $\mathcal{L}^1V(\pi,1) > \mathcal{L}^0V(\pi,1).$ In this case the optimal action is to play the arm. 
 \item For $y =1$ and   $\forall \pi \in [0,1],$ $\mathcal{L}^1V(\pi,1) <
   \mathcal{L}^0V(\pi,1).$ in this case not playing the arm is always optimal.
 \item There exists a $\pi_T \in (0,1),$ such that,
   $\mathcal{L}^1V(\pi,1) > \mathcal{L}^0V(\pi,1)$ for all $\pi<\pi_T,$ and $\mathcal{L}^1V(\pi,1) <\mathcal{L}^0V(\pi,1)$ for for all $\pi>\pi_T.$ Here,  $\pi_T$ is a threshold at which both  actions are optimal and obtain same value from both the actions.
 \end{enumerate}
 \label{def:threhold}
\end{definition}
To claim the existence of threshold type policy result we prove following structural properties of the value functions. { Using these properties, we will show that the optimal value function is submodular. The optimal threshold policy result follows from submodularity.} 

\begin{lemma}
{
For both stochastic and semi-deterministic availability,} 
\begin{enumerate}
\item value functions $V(\pi,y),$ and $\mathcal{L}^{a}_{w}V(\pi,y)$ are convex in $\pi$ for $a\in \mathcal{A}_y,$  $y\in\{0,1\}.$ 

\item value functions  $V(\pi,y)$ and $\mathcal{L}_{w}^{a}V(\pi,y)$ are convex in $w$ for all $\pi\in[0,1],$  $a\in \mathcal{A}_y,$ $y\in\{0,1\}.$
\end{enumerate}
\label{lemma:valfuncprops}
\end{lemma}
The proof is given in Appendix~\ref{proof:valfunc_convexity}.%

{ We note that convexity of value function is not enough to show threshold policy. This is because the belief update for played arm is non-linear in current belief. Even with some structural assumptions on transition probabilities, it is difficult prove submodularity. This is more clear from \cite[Lemma $2.1$ and Eqn.(4)]{Lovejoy87}, where submodularity and threshold behavior is proved when either playing or not playing action provides perfect state information. This is not true in our model. Hence we require an alternative proof technique. This is given in the following. }
{
\begin{remark}
\begin{itemize} 
\item We know about the continuity and convexity of value functions in $\pi$. We also know that value functions are absolutely continuous in $\pi.$ Further, value functions are Lipschitz in $\pi,$ this is because rewards are bounded and discounted with parameter $0<\beta <1.$ Hence, partial derivative of value function w.r.t. $\pi$ is bounded. Next, in Lemma~\ref{lemma:derivativewrt_pi_bound} , we derive a tight Lipschitz constant. This constant will be used in subsequent lemmas to prove submodularity and threshold policy result.
\item 
A tighter Lipschitz constant allows a wider range of transition probabilities for which threshold policy result can be proved analytically. A more relaxed Lipschitz constant gives a smaller range of transition probabilities for which the result is provable. We believe this is a technical limitation that does not allow us to  leverage the structure of the  problem to find out the smallest Lipschitz constant.
\end{itemize}
\end{remark} 
}

Now, we show that the partial derivative of the value function w.r.t. $\pi$ is bounded.  A tighter bound is derived under some conditions on state transition probabilities. {The bound is obtained under the assumption that $\theta$ is independent of $\pi.$ When $\theta$ is dependent on $\pi,$ we need  additional properties on the value function which are mentioned after the next Lemma.}


%
%

{
\begin{lemma}
	Given that $\theta^a_{y},y\in\{0,1\},a\in\mathcal{A}_y$ is independent of $\pi$ and for any of the following conditions,
	\begin{enumerate}
		\item ${0<p_{00}-p_{10}<\frac{1+b}{b+3c}},$
		\item $0<p_{10}-p_{00}<\frac{1+b}{b+c},$ 
	\end{enumerate}
	absolute values the derivatives of the action value functions are bounded, \textit{i.e.,}
	$\left|\frac{\partial \mathcal{L}^0 V(\pi,1)}{\partial \pi} \right|,$ $\left|\frac{\partial \mathcal{L}^1 V(\pi,1)}{\partial \pi}\right|$ and $\left|\frac{\partial V(\pi,1)}{\partial \pi}\right|$ are bounded by $\kappa c(\rho_1 - \rho_0),$ where $ \kappa = \frac{1}{(1-\beta|p_{00} - p_{10}|)} >1,$  $b = \frac{\eta_1-\eta_0}{\rho_1-\rho_0},$ $c = \max\left\lbrace 1, \frac{\eta_1-\eta_0}{\rho_1-\rho_0} \right\rbrace.$ 
	\label{lemma:derivativewrt_pi_bound}
\end{lemma}
}
{The proof is given in Appendix~\ref{proof:derivativewrt_pi_bound}.}
{
	\begin{remark} 
	\begin{itemize} 
	\item When $\rho_0 = \eta_0$ and $\rho_1 = \eta_1,$ the bound on the derivatives in Lemma~\ref{lemma:derivativewrt_pi_bound}  becomes $\kappa (\rho_1-\rho_0)$ under the conditions  ${0<p_{00}-p_{10}<1/2},$ or $0<p_{10}-p_{00}<1.$
	\item Also notice that when $\rho_0 = \eta_0$ and $\rho_1 = \eta_1,$ we can have a better Lipschitz constant with less restrictive conditions on transition probabilities. For example, let the value of ratios $b,c$ be $b = c = 1.1$. The conditions become $0<p_{00}-p_{10}<0.477$ or $0<p_{10}-p_{00}<0.954$.
\end{itemize} 
\end{remark} 
}

{ 
\begin{remark}
In Lemma~\ref{lemma:derivativewrt_pi_bound} we had assumed that $\theta^a_y$ is independent of $\pi.$  Instead, suppose $\theta^a_y(\pi)$ is a linear function of $\pi$ for given $a$ and $y.$ To obtain a bound on the partial derivative of the value functions w.r.t. $\pi,$ we need to bound $(V(\pi,1) - V(\pi,0))$ in terms of $\rho_0 - \rho_1$ (see eqn.\eqref{derivative1}). It is  difficult to derive a tight bound  because of the additional term $(V(\pi,1) - V(\pi,0)).$  We believe that one can have loose bound on $(V(\pi,1) - V(\pi,0))$ which may introduce a more stringent condition on the difference of transition probabilities and in turn a loose Lipschitz constant. Hence, we do not analyze this scenario here. 
\end{remark}
}


Let  $D(\pi) := \mathcal{L}^1 V(\pi,1)-\mathcal{L}^0V(\pi,1).$ It  gives the advantage of playing the arm in belief state $\pi$ when it is available.  The following lemma states that this advantage decreases as the belief  increases.
%



%
%
{
\begin{lemma}
	Given that $\theta^a_{y},y\in\{0,1\},a\in\mathcal{A}_y$ is independent of $\pi$ and for any of the following conditions,
	\begin{enumerate}
		\item ${0<p_{00}-p_{10}<\frac{b}{b+3c+1}},$
		\item  or $0<p_{10}-p_{00}<\frac{b}{b+c+1},$ 
	\end{enumerate} 
	the function $D(\pi)$ is decreasing in $\pi.$  
	\label{lemma:advantage}
\end{lemma}
}
{The proof is given in Appendix~\ref{proof:advantage}}.
\begin{remark}
Note that $\mathcal{L}^1 V(\pi,1)$ and $\mathcal{L}^0V(\pi,1)$ are convex in $\pi$.  Convexity of value functions and the preceding Lemma~\ref{lemma:advantage} suggest that $D(\pi)$ has at most one root in $\pi \in (0,1).$ 
\label{rem:dpiroot}
\end{remark}

{
We now state our main result, the optimal threshold policy result.
\begin{theorem}
\label{thm:opt-threshold-policy}
Constrained restless single armed bandits of stochastic and semi-deterministic availability types satisfying either  
\begin{enumerate} 
 \item ${0<p_{00}-p_{10}<\frac{b}{b+3c+1}}$ or
 \item $0<p_{10}-p_{00}<\frac{b}{b+c+1},$ 
\end{enumerate}
admit an optimal policy of threshold type.
\end{theorem}
}
\begin{IEEEproof}
 Suppose $\mathcal{L}^1V(\pi,1)>\mathcal{L}^0V(\pi,1),$ at $\pi = 0 .$ That is, playing the arm is advantageous than not playing it. 
 From Lemma \ref{lemma:advantage}, $D(\pi)$ can have at most one root in $[0,1].$\\
 Case 1) $D(\pi)$ has a root in $[0,1]:$ From Lemma \ref{lemma:advantage}, we know that this advantage decreases as $\pi$ increases. So, there exists a $\pi_T \in(0,1) : D(\pi) = \mathcal{L}^1V(\pi,1) < \mathcal{L}^0V(\pi,1)<0, \forall \pi > \pi_T .$ Hence the policy is of threshold type by definition \ref{def:threhold}.\\
 Case 2) $D(\pi)$ has no root in $(0,1):$ This means $D(\pi)>0,\pi \in (0,1).$ Hence the optimal policy always choose to play the arm and is threshold type by definition. 
 Similar arguments can be made when $\mathcal{L}^1V(\pi,1)<\mathcal{L}^0V(\pi,1),$ to claim the result.

\end{IEEEproof}


\subsection{Indexability}


{Index for a constrained restless single armed bandit in a given state is defined as the minimum amount of subsidy for which the value of not playing the arm  becomes greater than or equal to the value of playing the arm. As subsidy is provided for not playing the arm, a higher value of the index indicates greater gains from playing the arm. To use these indices in decision making  they need to be well defined. For this we need to first prove indexability of CRSABs.  In this section, we define indexability for an arm (CRSAB) and prove that it is indexable. To claim this we make use of the optimal threshold policy result. Hence, we assume same conditions on transition probabilities similar to those in Theorem~\ref{thm:opt-threshold-policy}.}


We now define indexability for a CRSAB and provide sufficient conditions. 

For a given subsidy $w,$ let $\mathcal{G}(w)$ be a set formed by members $(\pi,y)$ of perceived state space $S = [0,1]\times \{0,1\}$ for which not playing the arm when available is optimal. That is, 
{{ \begin{eqnarray*}
\mathcal{G}(w) := {\{ [0,1]\times\mathcal{A}_{0} \} \cup} {\{[0,1]\times\mathcal{A}_{1} : \mathcal{L}^1 V(\pi,1) \leq \mathcal{L}^0V(\pi,1)\}} . \nonumber
\end{eqnarray*} }}

\begin{definition}(Indexability)
 The arm is indexable if the set $\mathcal{G}(w)$ is increasing in $w
 \in \mathbb{R}.$
\end{definition}
 
Intuitively, indexability suggests that, if not-playing is the optimal choice for a  given subsidy $w,$ then it is also the optimal choice at higher values of subsidy  $w^{\prime} >w.$

{
\begin{remark}
\begin{itemize}
\item Action value functions $\mathcal{L}_{w}^{1}V(\pi,1)$ and $\mathcal{L}_{w}^{0}V(\pi,1)$ are non-decreasing and strictly increasing in subsidy $w,$ respectively. The proof of this straightforward, it uses the principle of mathematical induction. 
\item For a CRSAB, Theorem~\ref{thm:opt-threshold-policy} shows that there exists a threshold belief $\pi_T\in (0,1)$ at the optimal action switches from playing the arm to not playing as we cross over to its right from the left. This threshold is a function of subsidy $w.$ In the following, we will see that the threshold  moves left the segment as subsidy increases.
\end{itemize}
\end{remark}
}
To prove indexability of the arm, we use the following Lemma from \cite{Meshram18} and provide a sketch of the proof. 

\begin{lemma} 
Let 
\begin{equation*}
	\pi_{T}(w) = \inf  \{0\leq \pi \leq 1 : \mathcal{L}_{w}^1 V(\pi,1) \le \mathcal{L}_{w}^0 V(\pi,1)\} \in [0,1].
\end{equation*}  
If $\frac{\partial \mathcal{L}^1 V(\pi,1)}{\partial w}\left| _{\pi=\pi_{T}(w)} < \frac{\partial \mathcal{L}^0 V(\pi,1)}{\partial w}\right|_{\pi=\pi_{T}(w)},$
then $\pi_{T}(w)$ is a monotonically decreasing function of $w.$
\label{lemma:thresholddecreasesinpi}
\end{lemma}

\textit{Proof sketch:} This proof is by contradiction. Assume that thresholds $\pi_T(w) < \pi_T(w')$ for $w<w',$ under given `if' condition. By definition of threshold, $\mathcal{L}_{w}^1 V(\pi_T(w),1) = \mathcal{L}_{w}^0 V(\pi_T(w),1).$ For some $w' = w+\epsilon,$ $\epsilon\in (0,c),c<1,$ we have  $\mathcal{L}_{w'}^1 V(\pi_T(w'),1) \geq \mathcal{L}_{w'}^0 V(\pi_T(w'),1).$ This means   
$\frac{\partial \mathcal{L}^1 V(\pi,1)}{\partial w}\left| _{\pi=\pi_{T}(w)} > \frac{\partial \mathcal{L}^0 V(\pi,1)}{\partial w}\right|_{\pi=\pi_{T}(w)},$ which contradicts our assumption. \qed \\

Define $D(\pi,w) := \mathcal{L}_{w}^{1}V(\pi,1)-\mathcal{L}_{w}^{0}V(\pi,1).$

{
\begin{theorem}
A CRSAB with bounded subsidy $w\in [w_l,w_h]$ and discount parameter $\beta \in (0,1),$ is indexable under either of the following conditions.
\begin{enumerate} 
	\item  $0<p_{00}-p_{10}<\frac{b}{b+3c+1}$ or
	\item  $0<p_{10}-p_{00}<\frac{b}{b+c+1}.$ 
\end{enumerate}
\end{theorem} }
\begin{IEEEproof}
The proof proceeds in the following steps. (1) From Lemma~\ref{lemma:valfuncprops}, it can be seen that the functions $\mathcal{L}_{w}^{a}V(\pi,y)$ $a\in\mathcal{A}_y$ are convex and Lipschitz in $w.$ This implies that they are absolutely continuous. (2) It means $D(\pi,w)$ is absolutely continuous, which implies that it is differentiable w.r.t $w$ almost everywhere in the interval $[w_l,w_h],$ for all $\pi \in [0,1].$ (3) This implies that the threshold $\pi_T(w):= \{\pi\in[0,1]\mid D(\pi,w) = 0\}$ is absolutely continuous on $[w_l,w_h];$ hence, $\pi_T(w)$ is differentiable  w.r.t $w$ almost everywhere. (4) From Remark 2, $D(\pi,w)$ is decreasing in $w;$ hence, $\frac{\partial D}{\partial w}\leq 0$ almost everywhere in $[w_l,w_h].$ This implies $\frac{\partial \pi_T(w)}{\partial w}\leq 0.$ Now, using Lemma~\ref{lemma:thresholddecreasesinpi} we can say that $\pi_T(w)$ decreases with $w.$ This means as subsidy $w$ increases,  the set $\mathcal{G}(w)$ also increases. Hence, the arm is indexible.
\end{IEEEproof}
{
Indexability ensures a well defined index for an arm. Now we will be able use this index to define heuristic index based policies for solving the constrained restless multi-armed bandit problem. One such policy is the Whittle's index policy in which the arm with the highest index is played in each slot. Before proceeding to apply this policy to the CRMAB problem, we need an algorithm to compute Whittle's index. }

\subsection{Computing Whittle's index}
The index of a CRSAB is the minimum subsidy required to make not-playing the optimal action; it is defined below.
{ Note that closed form expressions for value functions are not available.}
 It is difficult to obtain a closed form expression for the index. We devise an algorithm for Whittle's index computation for CRSABs. The argument for convergence of this algorithm is based on stochastic approximation schemes. 


\begin{definition}(Whittle's index)
 For a given belief $\pi \in [0,1],$ Whittle's index $W(\pi)$ is the minimum subsidy for which, not playing the arm will be the optimal action. 
 \begin{equation*}
  W(\pi) = \inf\{w\in\mathbb{R}:\mathcal{L}_w^0V(\pi,1)\geq\mathcal{L}_w^1V(\pi,1) \}
 \end{equation*}
\end{definition}

An algorithm for computing Whittle's index. Algorithm~\ref{algo:WI} is based on two timescale stochastic approximation.  
\begin{algorithm}[h]
\KwIn{Reward values $\eta_0,\eta_1$, initial subsidy $w_0,$ tolerance $h,$ grid over belief space $ G([0,1])$.}
\KwOut{Whittle's index $W(\pi)$}
for $\pi\in G([0,1])$ \\
$w_t \gets w_0$;\\
\While {$|\mathcal{L}_{w_t}^1V(\pi,1)-\mathcal{L}_{w_t}^0V(\pi,1)|>h$ }
{\begin{align} 	
w_{t+1} = w_t + \alpha_t(\mathcal{L}_w^1V(\pi,1)-\mathcal{L}_w^0V(\pi,1)), 
\label{eqn:update_w_t} 
\end{align}
\begin{align*}
&t=t+1; \text{  compute } \mathcal{L}_w^0V(\pi,1), \mathcal{L}_w^1V(\pi,1).\nonumber 
\end{align*} 
}
\Return{$W(\pi,1) \gets w_t$}\;
\caption{{\sc WI} computes Whittle's index for CRSAB}
\label{algo:WI}
\end{algorithm}

The algorithm runs on two timescales; value iteration algorithm runs on the faster timescale, while subsidy $w$ is updated on the slower timescale. That is, the value function $\mathcal{L}_w^0V(\pi,1)$, $\mathcal{L}_w^1V(\pi,1)$ are updated on faster timescale, while the value of $w_t$ is updated along the slower one. In this algorithm, the algorithm on the faster timescale views $w_t$ as quasi-static, and  runs value iteration till convergence. Whenever $|\mathcal{L}_{w_t}^1V(\pi,1)-\mathcal{L}_{w_t}^0V(\pi,1)|>h,$ then $w_t$ is updated according to equation~\eqref{eqn:update_w_t}. Otherwise, subsidy or index  $W(\pi,1) = w_t.$ 

In stochastic approximation, two-timescale algorithms converge if the sequence $ \alpha_t $ is decreasing, $ \sum_t \alpha_t = \infty $ and $\sum_t \alpha_t^2 <\infty.$ This convergence is almost sure as shown in  Theorem $2$, \cite[Chapter 6]{Borkar08}. If $ \alpha_t $ is replaced with a tiny constant value $ \alpha, $ there is convergence with high probability; see \cite[Section 9.3]{Borkar08}.
%
%

A closed form index formula is not feasible for our model and the stochastic approximation algorithm is computationally expensive. This motivates us look for alternate heuristic policy without compromising on performance.

\section{Sample complexity results}
\label{sec:sample_complexity}
In the following we discuss sample complexity results for Whittle's index policy. The line of thought is as follows.
\begin{enumerate}
\item We used two-timescale stochastic approximation (TTSA) for Whittle's index computation in Algorithm~\ref{algo:WI}. The subsidy $w$ is updated along the slower timescale and value iteration is performed along the faster timescale.
\item One way to estimate the complexity of Algorithm~\ref{algo:WI} is by using available sample complexity results for one-timescale stochastic approximation (OTSA) (along the slower timescale) and value iteration (along the faster timescale). So the overall complexity would be the number of iterations for convergence of OTSA multiplied by number of iterations for convergence of value iteration.
\item However, the above approach might only give a loose estimate of the complexity of Algorithm~\ref{algo:WI}.
\item Hence, we discuss other possible approaches to estimate complexity such as the primal dual algorithm of discounted MDPs. 
\end{enumerate}

We now discuss some known results regarding the sample complexity of SA and value iteration.
Recently, there has been surge of interest in finite time analysis of two-timescale stochastic approximations~\cite{Dalal2018,Borkar18,Kaledin2020,Doan2021}. This analysis claims the existence of finite time $T^*$ after which the iterate (algorithm) remains in the $\epsilon$-neighborhood of the optimal solution,  with high probability. Concentration bounds are utilized to show this. 

There has been work on sample complexity of solving discounted MDPs with value iteration and Q-learning algorithms, \cite{Qu2020, Sidford2020}. Sample complexity results provide explicit value (expressions) of $T^*$ for which the iterate converges to $\epsilon$-optimal solution. These are stronger claims than finite time analysis, and are often difficult to obtain. No direct results are available on the sample complexity of TTSA algorithms. Even for OTSA in a general setting, few results are available till date \cite[Section $4.2$]{Borkar08},\cite[Section VI]{Karmakar2021}. Using the available literature, we comment on sample complexity of Whittle's index computation algorithm. 

We state all results for the constrained restless single-armed bandit model.  

Sample complexity of value iteration for MDP problems was given by \cite[Section $4.3$]{Littman95}. 
We note that our state space is the belief space $[0,1].$ We discretize this state space using a grid.  Let $G([0,1])$ denote a grid over the interval $[0,1],$ and $|G|$ denotes the number of grid points. 
In our value iteration algorithm, the number of computations required in each iteration is $O(2|G|^2).$ 
The maximum number of iterations required to find $\epsilon$-optimal policy is  
\begin{eqnarray*}
T^* \leq \frac{B + \log(1/\epsilon) + \log(1/(1-\beta)) +1}{1-\beta}. 
\end{eqnarray*}
The derivation can be found in \cite{Littman95}. Here, $B$ is in bits, which is the memory size in a linear program. The  number of iterations is polynomial in $|G|, B$ and $1/(1-\beta).$ The lower bound on $T^*$ is given as follows,
\begin{eqnarray*}
T^* \geq  \frac{1}{2} \frac{1}{1-\beta} \log \left( \frac{1}{\epsilon(1-\beta)} \right)
\end{eqnarray*}


The lower bound is summarized in the following Lemma. 
\begin{lemma}[Lower Bound]
The worst case complexity of computing value functions $\mathcal{L}^0_w V(\pi,1)$ and $\mathcal{L}^1_w V(\pi,1)$ $\forall \pi\in G([0,1])$ with an error tolerance $\epsilon$ is $\Omega(|G|^2T^*)$ with $T^* = \frac{1}{1-\beta} \log \left( \frac{1}{\epsilon(1-\beta)}\right).$ 
\end{lemma}

Here, the minimum number of iterations required for convergence to an $\epsilon-$optimal answer is given by $T^*.$ The number of computations needed for each iteration is at most $2|G|^2$ as there are two possible actions.

More recently, faster variants of value iteration algorithm have been proposed in \cite{Sidford2020}. For example, the high precision randomized value iteration, which provides bounds on sample complexity with high probability. This bound depends on the number of states, actions, discount parameter, $\epsilon$ (near optimal parameter) and $\delta$ (probability confidence). In the derivation of this bound,  concentration inequalities are used. We state the result \cite[Lemma $4.9$]{Sidford2020} for our single armed bandit without proof. 
For our case, this sample complexity for obtaining $\epsilon$ optimal value function with probability $(1- \delta )$ can be given as 
\begin{equation}
\bar{O}\left(\left( |G|^2+ \frac{|G|}{(1-\beta)^3} \right)\log \left(\frac{R_{max}}{\epsilon} \right) \log \left(\frac{1}{\delta} \right) \right),
\label{eqn:sample-complex-rand-value}
\end{equation}
where $R_{max}$ is the maximum possible reward over all states and actions and $\bar{O}$ is used to hide a  polylogarithmic factor in the input, i.e., $\bar{O}(f(x)) =O\left(f(x) \left(\log (f(x))\right)^{O(1)}\right).$ It is important to note that the sample complexity has polynomial dependence on $\frac{1}{(1-\beta)}.$  

For Whittle's index computation algorithm, sample complexity results are unknown and very challenging to determine from  finite time analysis of TTSA. Sample complexity results known for SA are not very informative (see \cite[Chapter $4.2$]{Borkar08} and \cite{Karmakar2021}) in the sense that they claim convergence to the optimal neighbourhood after a finite `large enough' number of iterations. This `large enough' number is however not known. 
For estimating the complexity of Algorithm~\ref{algo:WI}, we can utilize sample complexity results derived for constrained discounted Markov decision processes with primal-dual methods, \cite{JZhang21}.   

Primal dual algorithm is used in solving constrained optimization problem, \cite{Nocedal06}. Basically in this one would like to find a solution to Lagrangian relaxed problem which is unconstrained. It is required  to find optimal solution to both primal and dual variables; this boils down to finding the saddle point condition. In primal dual algorithm, the idea is to fix dual variables, i.e., Lagrangian multipliers, and solve the problem for primal variables. The optimal solution obtained for the primal is dependent on the dual variables. If we change the dual variable according to gradient ascent/descent method, then we get a new optimal primal solution. Thus, these primal-dual solutions are coupled iterations and can be analyzed using ideas of two timescale approach, where dual variables are updated on slower timescale (assumed as quasi-static) and the primal variables are updated on the faster timescale.  Such two timescale approach can be analyzed using TTSA algorithms, \cite[Chapter $5$]{Borkar08}.  Similar ideas are employed for constrained discounted MDPs in \cite{Borkar05} in case of actor-critic methods in constrained MDPs.  For the sake of clarity we describe TTSA algorithm from \cite[Chapter $6$]{Borkar08} below. 
\begin{eqnarray}
\zeta_{t+1} = \zeta_t + \kappa_t \left[ f(\zeta_t, w_t) + M^1_{t+1}\right], \\
w_{t+1} = w_t + \alpha_t \left[ g(\zeta_t, w_t) + M^2_{t+1}\right], 
\end{eqnarray}
where, $M^1_{t+1},$ and $M^2_{t+1}$ are martingale difference noise terms.
The $\{\kappa_t \}$ and $\{ \alpha_t \} $ are  stepsizes $t \geq 1$ and satisfy the following condition:  
\begin{eqnarray}
\sum_{t =1}^{\infty} \kappa_t = \sum_{t =1}^{\infty} \alpha_t = \infty, \\
\sum_{t=1}^{\infty} \left(\kappa_t^2 + \alpha_t^2 \right) < \infty, \\
\frac{\alpha_t}{\kappa_t} \rightarrow 0, \mbox{as $t \rightarrow \infty.$} 
\end{eqnarray}
The last condition on step sizes implies that $w_t$ moves on slower timescale than $\theta_t.$ The convergence analysis is given in \cite[Chapter $6$]{Borkar08}. This analysis is also valid when $\frac{\alpha_t}{\kappa_t} = \frac{\alpha}{\beta} = \epsilon_1 <<1,$  $\alpha_t = \alpha <1,$  $\kappa_t = \kappa < 1$ for $t \geq 1$ but convergence guarantees are in probabilistic sense instead of almost sure convergence. 


Now coming back to sample complexity results, we would like to state a result similar to \cite{JZhang21}, which uses two-timescale approach in discounted MDP and make use of primal-dual stochastic algorithm. In \cite[Theorem $1$]{JZhang21}, the duality gap is expressed as function of $T,$ state $S,$ action $A$ and discount parameter and this in turn describes the convergence rate. In their algorithm, the step sizes are chosen such that two timescale behavior holds in their setting. Using these concepts, they derive sample complexity result in terms of number of states, number of actions, discount parameter and desired accuracy. We state their result in our framework using similar step sizes and conjecture that this is the optimal sample complexity for Whittle index computation with a single-arm restless bandit. Suppose the step size $\alpha_t = \alpha$ for update rule $w_t$ is set to $\frac{1-\beta}{\widetilde{C}}\sqrt{\frac{\log(2|G|)}{2T|G|}},$  and value iteration algorithm in index computation algorithm is updated at timescale $\sqrt{\frac{|G|}{T}} \widetilde{C}$ for all $t.$
Note that this choice of step sizes  satisfy two-timescale conditions discussed before. In fact the ratio of these is $\frac{(1-\beta)}{\widetilde{C}^2}\sqrt{\frac{\log (2|G|)}{2|G|^2}} << 1,$ where $\widetilde{C}>1.$  Then, we expect the following result. 
\begin{result}[Conjecture on sample complexity of index computation]
For Whittle index computation algorithm~\ref{algo:WI}, stepsize of $w_t$ update rule is $\frac{1-\beta}{\widetilde{C}}\sqrt{\frac{\log(2|G|)}{2T|G|}},$ and value iteration is updated at faster timescale with stepsize $\sqrt{\frac{|G|}{T}} \widetilde{C}.$  Let $\widehat{w}$ be the output of the Algorithm~\ref{algo:WI}, and $w^*$ be the true value of index at which the value functions for both actions are equal. Then, we have 
\begin{eqnarray}
\vert V_{\widetilde{w}}(\pi,y) - V_{w^*}(\pi,y)  \vert \leq O\left( \sqrt{\frac{2|G| \log (2 |G|)}{T}} \frac{\widetilde{C}}{(1-\beta)^2} \right ).
\end{eqnarray}
To guarantee $\vert V_{\widetilde{w}}(\pi,y) - V_{w^*}(\pi,y)  \vert \leq \epsilon,$ the optimal number of  sample  required is 
\begin{eqnarray}
T = O \left( \frac{2|G| \log(2|G|) \widetilde{C}^2}{(1-\beta)^4 \epsilon^2} \right).
\label{eqn:sample-complex-index-compute}
\end{eqnarray}
\end{result}
Here, $V_{w}(\pi,y)$ denotes the optimal value function (defined in \eqref{eqn: Valfunc_def}) with subsidy $w.$
We do not provide a proof here; we believe this result will hold true from the analysis of \cite{JZhang21} and our result will require similar type of analysis. 
\begin{remark} 
From comparison of sample complexity result in Eqn.~\eqref{eqn:sample-complex-index-compute} with sample complexity of randomized value iteration algorithm in Eqn.~\eqref{eqn:sample-complex-rand-value}, we observe that there is increase in number of samples needed by factor of $\frac{1}{(1-\beta) \epsilon^2}. $ 
\end{remark}

\section{Online Rollout Policy}
\label{sec:rollout}
{ We now present a simulation based approach referred to as online roll-out policy. In the Whittle's approach the constraint $\sum_{i=1}^{N}a_i(t) = M$ is first relaxed to a discounted form $\sum_{t=1}^{\infty} \sum_{i=1}^{N}\beta ^{t-1} a_i(t) = \frac{M}{1-\beta},$ then Lagrangian relaxation method is applied. Instead, we directly employ a simulation based look-ahead approach. We call this the rollout policy as many trajectories are `rolled out' using a simulator and the value of each action is estimated based on the cumulative reward along these trajectories. The details are given below.

	Trajectories of length $H$ are generated using a fixed base policy, say, $\phi,$ which might choose arms according to a deterministic rule (say, myopic decision) at each step. The information obtained from a trajectory is 
	\begin{eqnarray} 
	\{ \pi_{n}(h,l), y_{n}(h,l), b_n^{\phi}(h,l), R_{n}^{\phi}(h,l) \}_{n=1,h=1}^{ N, H}
	\end{eqnarray} 
	under policy $\phi.$ Here, $l$ denotes a trajectory, $h$ denotes time step and $\pi_{n} (h,l)$ denotes the belief about arm $n.$ The action of playing or not playing arm $n$ at step $h$ in trajectory $l$ is denoted by $b_n^{\phi}(h,l) \in \{0,1\}.$ Reward obtained from arm $n$ is $R_{n}^{\phi}(h,l)$ and the availability of arm $n$ is denoted by $y_{n}(h,l) \in \{0,1\}.$ } Recall that the play of an arm depends on availability of that arm. 

    { We now describe the rollout policy for $M=1.$ 	
	%
	We compute the value estimate for trajectory $l$ with starting belief $\bm{\pi} = (\pi_{1}, \cdots, \pi_N),$ availability $\bm{y} = (y_1, y_2, \cdots, y_n),$  and initial action $\xi\in \{1,2, \cdots, N\}.$  Here, $\xi(h,l) = n$ means arm $n$ is played at step $h$ in trajectory $l$, so $b_n^{\phi}(h,l) = 1,$ and  $b_i^{\phi}(h,l) = 0$ for $\forall i \neq n.$
	The value estimate for initial action $\xi$ along trajectory $l$ is given by 
	\begin{eqnarray*}
		Q_{l}^{\phi}( \bm{\pi}, \bm{y}, \xi ) &=& \sum_{h=1}^{H} \beta^{h-1} \sum_{n=1}^{N} R_{n}^{\phi}(h,l)  \\
		&=& \sum_{h=1}^{H} \beta^{h-1} \sum_{n=1}^{N} r_n(\pi(h,l), y_n^{\phi}(h,l), b^{\phi}_n(h,l)).
	\end{eqnarray*} 
	Then, averaging over $L$ trajectories the value estimate for action $\xi$ in state $\pi$    under policy $\phi$ is 
	\begin{eqnarray*}
		\widetilde{Q}_{H,L}^{\phi}({\bm{\pi},  \bm{y}, \xi}) = \frac{1}{L}\sum_{l=1}^{L}  Q_{l}^{\phi}(\bm{\pi}, \bm{y}, \xi).
	\end{eqnarray*} 
	Here, the base policy $\phi$ is myopic (greedy), it chooses the arm with the highest immediate reward, along each trajectory. Now we perform one step policy improvement, and the optimal action is selected as, 
	\begin{align}
	j^*(\bm{\pi}, \bm{y}) = \arg \max_{1 \leq j \leq N} \left[ \widetilde{r}(\bm{\pi}, \bm{y}, \xi = j) + \beta \widetilde{Q}_{H,L}^{\phi}(\bm{\pi},\bm{y},\xi=j) \right]. 
	\label{eqn:policy-improv}
	\end{align}
	Here, $\widetilde{r}(\bm{\pi}, \bm{y}, \xi ) = \sum_{n=1}^{N} r_n(\pi_n, y_n, b_n ).$
	
	In each time slot $t$ with belief  $\bm{\pi}(t)$ and availability $\bm{y}(t)$, online roll-out policy plays the arm $j^*(\bm{\pi}(t),\bm{y}(t))$ obtained according to \eqref{eqn:policy-improv}.  
	
	The detail discussion on rollout policy for MDP and restless bandits with complex action space is given in \cite{Meshram20}. In \cite{Meshram21a, Meshram21b}, roll-out policy is extended to partially observable restless multi-state restless bandits. 
}
\subsection{Playing multiple arms using online roll-out policy}
{   
    Our discussion above consider the case $M=1,$ that is, only one arm is played in each time slot. In particular this is assumed while  employing the base policy $\phi.$ 
	When a decision maker plays more than one arm  per slot, employing a base policy with future look-ahead is non-trivial. This is due to the large number of possible combinations of $M$ out of $N$ available arms, i.e., $\binom{N}{M}$. Since the rollout policy depends on future look-ahead actions, it can be computationally expensive to implement as each time step we need to choose from $\binom{N}{M}.$ 
	We reduce these computations for base policy $\phi$ by employing a myopic rule in look-ahead approach, where we select $M$ arms with highest immediate rewards while computing value estimates of trajectories. } 

{
In this case, $\sum_{n=1}^{N}b_{n}^{\phi}(h,l) = M,$ with $M>1.$ The set of arms played at step $h$ in trajectory $l$ is  $\bm{\xi}(h,l) \subset \mathcal{N} = \{1,2, \cdots, N \},$ with $\vert \bm{\xi}(h,l) \vert = M.$  Here, $b^{\phi}_n({h,l})=1$ if $n\in\bm{\xi}(h,l).$
The base policy $\phi$ uses myopic decision rule and the one step policy improvement is given by 
\begin{eqnarray}
\bm{j}^*(\bm{\pi}, \bm{y}) = \arg \max_{\bm{\xi} \subset \mathcal{N}} \left[ \widetilde{r}(\bm{\pi}, \bm{y}, {\xi}) + \beta \widetilde{Q}_{H,L}^{\phi}(\bm{\pi}, \bm{y}, \bm{\xi}) \right]. 
\label{eqn:policy-improv-2}
\end{eqnarray}
Here, $\widetilde{r}(\bm{\pi},\bm{y} ,\bm{\xi}) = \sum_{n=1}^{N} r_n(\pi_n, y_n, b_n).$ 
The computation of $\widetilde{Q}_{H,L}^{\phi}(\bm{\pi},\bm{y},\bm{\xi})$ is similar to the preceding discussion. 
At time $t$ with belief $\bm{\pi}(t)$ and availability $\bm{y}(t),$  
rollout policy plays the subset of arms $\bm{j}^*$ obtained according to \eqref{eqn:policy-improv-2}. A more detailed discussion on rollout policy can be found in \cite{Meshram20}.
}




\subsection{Computational complexity}
We now present the computational complexity of online rollout policy. As rollout policy is a heuristic (lookahead)  policy  which does not require convergence analysis, and we only present its computational complexity.

WI computation is done offline where we compute and store the index values for each element on the grid $G,$ for all arms ($N$). 
During online implementation, when a belief state $[\pi_1,...,\pi_N]$ is observed, the corresponding index values are drawn from the stored data. On the other hand, online rollout policy is implemented online and its  computational complexity is stated in the following Lemma. 

\begin{lemma}
The online rollout policy has a worst case complexity of $O(|A|(HL+2)T)$ for number of iterations $T,$ when the base policy is myopic. Here $|A|$ is the number of possible actions in each iteration. 
\end{lemma}
\begin{IEEEproof}
\begin{itemize}
    \item Case $M=1$ (Only one arm is played): For each iteration, we need to compute the value estimates $\{\widetilde{Q}_{H,L}^{\phi}({\bm{\pi},  \bm{y}, \xi})\}$ for $N$ possible initial actions (arms). This takes $O(NHL)$ computations as there are $L$ trajectories of horizon (look ahead) length $H$ for each of the $N$ initial actions. For policy improvement step in Eqn~\ref{eqn:policy-improv}, it takes another $O(2N)$ computations. Thus total computation complexity in each iteration is $O(2N + NHL) = O(N(HL+2)).$ For $T$ time steps, the computational  complexity is $O(N(HL+2)T).$
    \item Case $M>1$ (Multiple arms are played): Here the number of possible actions is $\binom{N}{M}$ and  $\binom{N}{M} \approx O(N^M),$ which is a polynomial in $N$ for fixed $M.$ Thus the complexity would be very high if all the possible actions are considered.  The value estimates are computed only for some $|A|$ initial actions, $N\leq |A|< \binom{N}{M}.$  The computations required for value estimate $\widetilde{Q}_{H,L}^{\phi}(\bm{\pi}, \bm{y}, \bm{\xi})$ per iteration is $A H L.$ As there are $A$ number of  subsets considered in Eqn.~\ref{eqn:policy-improv-2}, the computations needed for policy improvement steps are at most $2|A|.$ So, the per-iteration computation complexity is $|A|HL + 2|A|.$ 
    Hence for $T$ time steps the computational complexity would be $O(|A|(HL+2)T).$  
\end{itemize}
\end{IEEEproof}

\begin{remark}
Note that computation complexity of online rollout policy in each iteration depends linearly on lookahead horizon length $H,$ number of trajectories $L$ and number of arms $N,$ in the case of where one arm is played. The offline computation of index (sample) complexity depends on the polynomial $\frac{1}{(1-\beta)^4},$ and is linear in number of arms $N.$
\end{remark}

\section{Bounds on optimal value functions}
\label{sec:bounds}
{In the previous sections we studied Whittle's index and online rollout policies as solutions to the CRMAB problem. Both these are heuristic policies which are  not necessarily optimal. The difference between the optimal value function and the value generated by a policy would be the absolute measure of goodness of a policy. However, it is hard to compute the optimal  value function of the original CRMAB problem over a polymatroid belief space. Hence, an upper bound on optimal value functions is computed to provide an estimate of the difference.  }


In this section we shall derive upper bounds on the optimal value function of a CRMAB. First, we shall compare its value function to that of a RMAB (unconstrained). In the following discussion we use the terms `unconstrained restless bandits' and `restless bandits' interchangeably.

\subsection{Relation between value functions of RMAB and CRMAB}
Let $U(\pi)$ be the value function of an restless single armed bandit which is always available. $U(\pi)$ is the solution of the following dynamic program.
\begin{align}
& U(\pi) = \max\{U_S(\pi),U_{NS}(\pi)\},\\
& U_S(\pi) = \eta(\pi) + \beta\bigg[ \rho(\pi)U(\Gamma_{1}(\pi)) + (1-\rho(\pi))U(\Gamma_{0}(\pi)) \bigg] \nonumber\\
& U_{NS}(\pi) = w + \beta U(\gamma^0_1(\pi)) .\nonumber
\end{align}
The following Lemma states that for the same Markov chain parameters, the optimal value of the restless single armed bandit is greater than that of constrained restless single armed bandit. 

{The following lemma states  that, when the arms of an RMAB are constrained the value generated by the optimal policy is decreased.}
\begin{lemma}
\label{lemma:unconst}
 For any given set of parameters $p_{00},$ $p_{10},$ $\rho_0,$ $\rho_1,$ $\eta_0,$ $\eta_1,$ $w,$ each of the following statements is true.
\begin{enumerate}
\item For belief update rules $\gamma^0_1(\pi) = \pi p_{00} + (1-\pi)p_{10}$ and $\gamma^0_0(\pi) = q,$ the inequality $U(\pi)\geq V(\pi,y),$ holds $\forall \pi\in \Pi_\Gamma ,y\in\{0,1\},$ where
{\footnotesize{ 
\begin{dmath}
\label{eqn:PiGamma}
{\Pi_\Gamma = \{\pi\in[0,1]|U_{NS}(q)\leq U_{NS}(\Gamma_{1}(\pi)), U_{NS}(\pi), U_{NS}(\Gamma_{0}(\pi)) \}}
\end{dmath} }}
\item If belief update rule $\gamma^0_1(\pi) = \gamma^0_0(\pi) = \pi p_{00} + (1-\pi)p_{10},$ the inequality $U(\pi)\geq V(\pi,y),$ holds $\forall \pi\in[0,1],y\in\{0,1\}.$
\end{enumerate} 
\end{lemma}
{The proof is straight forward, through induction.} 
%
%
\subsection{Bounds on value functions}
We shall now derive an upper bound on the value function of the constrained bandit.
The Lagrangian relaxation provides an upper bound on the value function of the original problem. This has been studied for weakly couple Markov decision processes by \cite{Hawkins03} and \cite{Adelman08}. {However, the applicability of this result for constrained availability case is not obvious. We extend this idea for CRMABs and provide proof for the generalized case of partial observability and constrained availability.} 

Let us now look at the constrained multi-armed bandit problem as a set of $N$ single armed bandits. We will be slightly abusing the notation in order to keep the mathematical expressions simpler; any change in notation is mentioned.

The CRMAB problem can be described as the following dynamic program. Given belief vector $\bm{\pi}\in [0,1]^N$ and availability vector $\bm{y}\in\{0,1\}^N,$ find $J(\bm{\pi},\bm{y})$ satisfying 

\begin{dmath}
J(\bm{\pi},\bm{y}) = \max\limits_{\bm{a}\in \mathcal{A}_{\bm{y}}} \bigg\lbrace R(\bm{\pi},\bm{y},\bm{a}) +  \beta\sum\limits_{\bm{o}\in{S_{\bm{o}}},\bm{y'}\in S_{\bm{y}}}\prob{\bm{o},\bm{y'}|\bm{\pi},\bm{y},\bm{a}} J(\bm{\Gamma^o}(\bm{\pi}),\bm{y'}) \bigg\rbrace \\
{s.t.\text{ } \lVert \bm{a}\rVert_1 = M, \text{  }\mathcal{A}_{\bm{y}} := \mathcal{A}_{y_1}\times \mathcal{A}_{y_2}\times ...\times\mathcal{A}_{y_n}.}
\label{eqn:crmab-wcpomdp}
\end{dmath} 
Here, $\bm{\Gamma^o}$ is the belief (vector) update rule for observation vector $\bm{o}.$ So, $\bm{\Gamma}^{o_n}$ is the belief update rule based on observation $o_n$ for arm $n.$ And $\bm{\Gamma^o}:S_{\bm{o}}\mapsto [0,1]^N.$  Here, $S_{\bm{o}}$ is the observation set with the set of all possible observation vectors. $S_{o_n}$ is the set of possible observations for arm $n.$ An observation vector $\bm{o}$ also contains some `no observation' elements corresponding to the unplayed arms.

The Lagrangian relaxed dynamic program of the above optimization problem is written as  
\begin{dmath}
\label{eqn:crmab-Lag_relaxed}
J^{\lambda}(\bm{\pi},\bm{y}) = \max\limits_{\bm{a}\in \mathcal{A}_{\bm{y}}} \bigg\lbrace R(\bm{\pi},\bm{y},\bm{a}) + \lambda [M-\lVert \bm{a} \rVert_1] \nonumber\\+	\beta\sum\limits_{\bm{o}\in{S_{\bm{o}}},\bm{y'}\in S_{\bm{y}}}\prob{\bm{o},\bm{y'}|\bm{\pi},\bm{y},\bm{a}} J^{\lambda}(\bm{\Gamma^o}(\bm{\pi}),\bm{y'}) \bigg\rbrace, \\ 
{\lambda \geq 0.}
\end{dmath} 
The following Lemma states that the Lagrange relaxed value function of CRMAB can be written as a linear combination of value functions of $N$ constrained single armed bandits.
\begin{lemma}
\label{lemma:Lagrangian_decouple}
\begin{dmath}
J^{\lambda}(\bm{\pi},\bm{y}) = \frac{M\lambda}{1-\beta} + \sum\limits_{n=1}^{N} J^{\lambda}(\pi_n,y_n),
\label{eqn:lemma-Lagrangian_decouple}
\end{dmath}
\vspace*{-0.2 cm}
where,
\vspace*{-0.2 cm}
{{
\begin{dmath*}
J^{\lambda}(\pi_n,y_n) = \max\limits_{a_n\in \mathcal{A}_{{y_n}}}\bigg\lbrace r_n(\pi_n,y_n,a_n)-\lambda a_n + \beta\sum\limits_{o_n\in S_{o_n}}\prob{o_n|\pi_n,y_n,a_n}\left[\theta^{a_n}_{y_n}J^{\lambda}(\Gamma^{o_n}(\pi_n),1) + (1-\theta^{a_n}_{y_n})J^{\lambda}(\Gamma^{o_n}(\pi_n),0) \right] \bigg\rbrace.
\label{eqn:lemma-decoupled_singlearmfunc}
\end{dmath*} }}
\end{lemma}
The proof in given in Appendix~\ref{proof:Lagrangian_decouple}.
%
The Lagrangian relaxed value function $U^{\lambda}(\bm{\pi})$ for an RMAB is given as follows (in \cite{Kaza19}). This provides an upper bound on Whittle's index policy for RMAB.
\begin{lemma}
\begin{dmath}
U^{\lambda}(\bm{\pi}) = \frac{M\lambda}{1-\beta} + \sum\limits_{n=1}^{N} U^{\lambda}(\pi_n),
\end{dmath} 
{{
\begin{dmath}
U^{\lambda}(\pi_n) = \max\limits_{a_n\in\{0,1\}}\bigg\lbrace r_n(\pi_n,a_n)-\lambda a_n + \beta\sum\limits_{o_n\in S_{o_n}}\prob{o_n|\pi_n,a_n} U^{\lambda}(\Gamma^{o_n}(\pi_n)) \bigg\rbrace.
\end{dmath} }}
\end{lemma}

The following corollary based on Lemma~\ref{lemma:unconst} and Lemma~\ref{lemma:Lagrangian_decouple} states that for the same set of state transition probabilities and rewards, the Lagrange relaxed value function of RMAB is greater than that of CRMAB. This means, an upper bound on value can be computed using either of the functions.
\begin{corollary}
The inequality $J^{\lambda}(\bm{\pi},\bm{y}) \leq U^{\lambda}(\bm{\pi})$ holds for each of the following cases. 
\begin{enumerate}
\item $\gamma_0^0(\pi) = q,$ $\gamma^0_1(\pi) = \pi p_{00} + (1-\pi)p_{10},$ $\forall \pi\in\Pi_\Gamma,y\in\{0,1\},$ 
\item $\gamma_0^0(\pi) = \gamma^0_1(\pi) = \pi p_{00} + (1-\pi)p_{10},$ $\forall \pi\in[0,1],y\in\{0,1\},$
{\footnotesize{ 
\begin{dmath*}
{\Pi_\Gamma = \{\pi\in[0,1]|U_{NS}(q)\leq U_{NS}(\Gamma_{1}(\pi)), U_{NS}(\pi), U_{NS}(\Gamma_{0}(\pi)) \}.}
\end{dmath*} }} 
\end{enumerate}
\end{corollary}
\begin{IEEEproof}
From Lemma~\ref{lemma:unconst} we know that the value functions of constrained restless single armed bandits are upper bounded by those of restless single armed bandits. It follows that their summation as given in Lemma~\ref{lemma:Lagrangian_decouple} is also similarly bounded.
\end{IEEEproof}
%

\section{Numerical Experiments}
\label{sec:simulation}
In this section we consider different parametric scenarios and evaluate the  performance of Whittle's index policy (WI), online rollout policy, modified Whittle's index policy (MWI) and myopic policy (MP) in terms of their value (discounted cumulative reward). {The impact of parameters such as the number of arms ($N$), number of played arms per slot ($M$), number of always available arms ($K$) and the reward structure is studied.} 

{Whittle's index policy plays the arms with  $M$ highest values of $(Y_n(t)W_n(\pi_n(t))).$  Myopic policy chooses the arms with $M$ highest expected immediate rewards, i.e., it considers the expression $Y_n(t)(\pi_n(t)\eta_{n,0}+(1-\pi_n(t))\eta_{n,1})$ as index for arm $n$.}
Modified Whittle index (MWI) is a less complex alternative to Whittle's index considered in \cite{Brown17,Kaza19}. However, its performance is found to be highly sensitive to problem parameters, in case of RMABs \cite{Kaza19}.
 It is defined for MDPs with finite horizon. The value of MWI at time $t$ is given as    
$m_t(\pi) = \mathcal{L}_{m_{t+1}}^1V(\pi,1)-\mathcal{L}_{m_{t+1}}^0V(\pi,1).$ 
 
In the following numerical examples, policies are evaluated for different bandit instances (a parameter set is called an instance). A bandit instance is specified by giving the values of 1) number of arms $N,$ 2) state transition probabilities of arms $p^n_{ij}(y,a),$ 3) availability probabilities of arms $\theta^a_n(y),$ 4) reward structure $\eta_{n,i},$ 5) success probabilities $\rho_n(i).$
For each bandit instance, the value function of each policy is computed, and averaged over numerous sample sequences of states and arm availability.

We now present the results of three experiments which will provide insight into the performance of various policies. {Discount factor $\beta$ is $0.99$  for all experiments.  Experiment-$0$ considers $10$-armed bandit instances and compares the performances of various policies in case of constrained and unconstrained availability.} Experiments $1$ $\&$ $2$ consider a $15$-armed bandit instance with same transition matrices and rewards, for stochastic and semi-deterministic availability models, respectively. 

{
\subsubsection{Experiment $0$ - Perfect observability for played arms}
We consider $10$-armed restless bandit instances, i.e. $N=10$. In this experiment we assume that exact state is observed for played arms, i.e. $\rho_0=0,\rho_1=1.$ The parameter set for stochastic availability model is given in Table.~\ref{Exp0_parameters}. We also present simulations for the semi-deterministic model. We use same parameters as in previous model where ever applicable, and chose $T_0 =3$ or $T_0=5$. For rollout policy we use $H=3$ and $L=100.$ 
A comparison of discounted rewards generated by Whittle's index policy and myopic policy for stochastic and semi-deterministic availability models is given in Table~\ref{table:Exp-0-results} and Tables~\ref{table:Exp-0-results-semid},\ref{table:Exp-0-results-semid-T0-5}, respectively. We observe that the rollout policy performs the best with Whittle's index policy being the close second. Also notice their closeness to the upper bound. For $M=1,$ both of them can be practically considered optimal. As $M$ increases they move away from the bound. Also notice that the performance of myopic policy gets closer to rollout and WI as $M$ increases. This might be due to the inherent sub-optimality of assigning an index value to each action using heuristic approaches (such as myopic, Whittle's index or rollout). Increase in $M$ increases the number of possible actions which accentuates the sub-optimality of these heuristics.

%
\begin{table}[h]
\centering
{
\caption{Experiment 0: Parameter set $(N=10,K=5)$}
\label{Exp0_parameters}
\begin{tabular}{|c|c|c|c|c|c|c|c|}
\hline
 Arm & $[\theta_1^1, \theta^0_1, \theta^0_0]$    & $\rho_0$  &$\rho_1$& $\eta_0$& $\eta_1$ & $p_{00}$ & $p_{10}$ \\ \hline
 1 	 & $[1, 1, 1]$ 		   & $0$    & $1$ 	&   $0$   &  $0.9$   & $0.5$ 	& $0.41$\\ \hline
 2 	 & $[0.3, 0.75, 0.8]$  & $0$  	& $1$ 	&   $0$	  &  $0.97$  & $0.45$	& $0.4$ \\ \hline
 3 	 & $[1, 1, 1]$  & $0$  & $1$ 	& $0$	&  $0.82$ &  $0.45$  & $0.35$\\ \hline
 4 	 & $[0.95, 0.9, 0.85]$ & $0$ 	& $1$ 	&   $0$	  &  $0.85$	 & $0.78$ 	& $0.15$\\ \hline
 5 	 & $[1, 1, 1]$ & $0$  & $1$ 	&   $0$	  &  $0.65$  & $0.6$ 	& $0.55$\\ \hline
 6 	 & $[1, 1, 1]$ & $0$  & $1$ 	&   $0$   &  $0.72$	 & $0.6$ 	& $0.5$\\ \hline
 7 	 & $[1, 1, 1]$ & $0$  & $1$ 	&   $0$   &  $0.75$	 & $0.7$ 	& $0.5$\\ \hline
 8 	 & $[0.8, 0.7, 0.6]$& $0$  & $1$ 	&   $0$   &  $0.45$	 & $0.7$ 	& $0.6$\\ \hline
 9 	 & $[0.9, 0.85, 0.95]$ & $0$  & $1$ 	&   $0$   &  $0.75$	 & $0.4$ 	& $0.3$\\ \hline
 10  & $[0.95, 0.9, 0.9]$ & $0$  & $1$ 	&   $0$   &  $0.72$	 & $0.45$ 	& $0.25$\\ \hline
\end{tabular}
}
\end{table}
\begin{table}[!h]
	\centering
	\caption{Experiment $0$ - Unconstrained availability and stochastic availability: Discounted Cumulative Rewards from WI and Myopic policy, with random initial belief.}
	\label{table:Exp-0-results}
	\begin{tabular}{|c|c|c|c|c|c|}
		\hline
		  Availability & Arms & $L_b$& WI      & Myopic  & Rollout\\
		     		 & played $(M)$& 	 &		   &	(MP)	 & policy \\ \hline
		  all 		 & $1$ & $62.55$ & $61.1$  & $56.2$  & $62.5 $\\ \hline 
		  stochastic & $1$ & $61.8$  & $59.11$ & $ 55$   & $ 61.46$\\ \hline 
		  stochastic & $2$ & $117.8$ & $111.6$ & $ 108$  & $113.9 $\\ \hline
		  stochastic & $3$ & $173.8$ & $162.9$ & $ 160.5$ & $ 166.1$\\ \hline
		  stochastic & $4$ & $229.8$ & $211.3$ & $ 210.2$ & $214 $\\ \hline
	\end{tabular}
\end{table}
}
\begin{table}[!h]
	\centering
	\caption{Experiment $0$ - semi-deterministic availability ($T_0=3$): Discounted Cumulative Rewards from WI and Myopic policy, with random initial belief.}
	\label{table:Exp-0-results-semid}
	\begin{tabular}{|c|c|c|c|c|c|}
		\hline
		Availability	   & Arms  		& $L_b$& WI 	 	& MP	   & Rollout  \\
					 	   & played& 	  &  		&	 	   & policy	\\ \hline 
		semi-deterministic & $1$ & $61.6$ & $58.28$ & $ 54.65$ & $60.37 $	\\ \hline 
		semi-deterministic & $2$ & $114.6$& $110.5$ & $ 107.5$ & $113 $	\\ \hline
		semi-deterministic & $3$ & $167.6$& $160.8$ & $ 159$   & $ 163.1$	\\ \hline
		semi-deterministic & $4$ & $220.6$& $206.9$ & $206$    & $208.1 $		\\ \hline
	\end{tabular} 
\end{table}
\begin{table}[!h]
	\centering
	\caption{Experiment $0$ - semi-deterministic availability ($T_0=5$): Discounted Cumulative Rewards from WI and Myopic policy, with random initial belief.}
	\label{table:Exp-0-results-semid-T0-5}
	\begin{tabular}{|c|c|c|c|c|c|}
		\hline
		Availability & Arms	  & $L_b$	  &   WI  & MP 		& Rollout \\
			 		 & played &  		  &	 	  &		 	& policy  \\ \hline 
		semi-deterministic & $1$ &$59.93$ & $56.5$& $ 53.8$ & $59.64 $		\\ \hline 
		semi-deterministic & $2$ &$112.93$&$109.6$& $ 107.1$& $ 111.7 $	\\ \hline
		semi-deterministic & $3$ &$165.93$& $159$ & $ 157.5$& $160.7  $	\\ \hline
		semi-deterministic & $4$ &$218.93$& $202$ & $ 202$ 	& $204.5 $		\\ \hline
	\end{tabular} 
\end{table}
\subsubsection{Experiment $1$ (stochastic availability) - Partially observable states}
We consider a $15$-armed bandit instance with stochastic availability model. The entire parameter set is given in \ref{Exp1_parameters}. The first five arms are always available while remaining are available according to action dependent probabilities.

\begin{table}[!h]
\centering
\caption{Experiment 1: Parameter set}
\label{Exp1_parameters}
\begin{tabular}{|c|c|c|c|c|c|c|c|}
\hline
 Arm & $[\theta_1^1, \theta^0_1, \theta^0_0]$    & $\rho_0$  &$\rho_1$& $\eta_0$& $\eta_1$ & $p_{00}$ & $p_{10}$ \\ \hline
 1 	 & $[1, 1, 1]$  & $0$ 	  	& $1$ 	& 	$0$   &  $0.65$  & $0.2$ 	& $0.5$\\ \hline
 2 	 & $[1, 1, 1]$  & $0$ 	   	& $1$ 	&   $0$	  &  $0.7$ 	 & $0.3$	& $0.5$ \\ \hline
 3 	 & $[1, 1, 1]$  & $0$ 	   	& $1$ 	&   $0$	  &  $0.75$	 & $0.4$ 	& $0.3$\\ \hline
 4 	 & $[1, 1, 1]$  & $0$ 	   	& $1$ 	&   $0$	  &  $0.8$	 & $0.5$ 	& $0.4$\\ \hline
 5 	 & $[1, 1, 1]$  & $0$	   	& $1$ 	&   $0$	  &  $0.85$  & $0.3$ 	& $0.3$\\ \hline
 6 	 & $[0.25, 0.8, 0.9]$& $0.1$ & $0.9$ &  $0.1$ &  $0.9$	 & $0.2$ 	& $0.8$\\ \hline
 7 	 & $[0.3, 0.9, 0.8]$ & $0.1$ & $0.7$ &  $0.1$ &  $0.7$	 & $0.3$ 	& $0.7$\\ \hline
 8 	 & $[0.4, 0.75, 0.7]$& $0.1$ & $0.8$ &  $0.1$ &  $0.8$	 & $0.4$ 	& $0.6$\\ \hline
 9 	 & $[0.5, 0.7, 0.4]$ & $0.2$ & $0.7$ &  $0.2$ &  $0.7$	 & $0.5$ 	& $0.5$\\ \hline
 10  & $[0.6, 0.8, 0.8]$ & $0.1$ & $0.7$ &  $0.1$ &  $0.7$	 & $0.3$ 	& $0.5$\\ \hline
 11  & $[0.7, 0.8, 0.7]$ & $0.2$ & $0.6$ &  $0.2$ &  $0.6$	 & $0.3$ 	& $0.3$\\ \hline
 12  & $[0.5, 0.5, 0.5]$ & $0.2$ & $0.8$ &  $0.2$ &  $0.8$	 & $0.6$ 	& $0.4$\\ \hline
 13  & $[0.8, 0.3, 0.4]$ & $0.3$ & $0.9$ &  $0.3$ &  $0.9$	 & $0.7$ 	& $0.3$\\ \hline
 14  & $[0.8, 0.4, 0.2]$ & $0.2$ & $0.9$ &  $0.2$ &  $0.9$	 & $0.8$ 	& $0.2$\\ \hline
 15  & $[0.7, 0.6, 0.6]$ & $0.3$ & $0.95$ & $0.3$ &  $0.95$	 & $0.9$ 	& $0.2$\\ \hline
\end{tabular}
\end{table}
 Table~\ref{Exp1_results} shows the discounted cumulative rewards achieved by various policies. 
 While rollout policy is still the best, 
 WI and myopic are very close behind.
 %
%
%
%
\begin{table}[!h]
\centering
\caption{Experiment $1$ Stochastic availability: Discounted Cumulative Rewards from various polices, with random initial belief.}
\label{Exp1_results}
\begin{tabular}{|c|c|c|c|c|c|c|}
\hline
	& $L_b$   & WI      & MWI    & Myopic & Rollout & Rollout  \\
	&  &  &  &  & ($H=3$) &  ($H=5$)   \\	\hline
 $Value$& $65.7$  & $64.7$  & $57.4$ & $64.3$ & $65$ & $65.4$ \\ \hline
\end{tabular}
\end{table}
%
%
%
%
%
%
%
\subsubsection{Experiment $2$ (semi-deterministic availability) - Partially observable states}
We again consider a $15$-armed bandit instance with semi-deterministic availability model. 
The parameters used for this experiment are same as in Experiment $1$, except for the availability parameters. Recall that semi-deterministic availability is characterized by parameters $[\theta_1^1, \theta^0_1, T_0].$ Here, $\theta_1^1, \theta^0_1$ are same as in Experiment $1$, and $T_0$ is chosen to be $3$ slots. The discounted cumulative rewards achieved by various policies are shown in tables~\ref{Exp2_results}. Again, the ordering on policy performance remains the same (as in Experiment 1), for $H=3$.  


%
\begin{table}[!h]
\centering
\caption{Experiment $2$ - Semi-deterministic availability: Discounted Cumulative Rewards from various polices, with random initial belief.}
\label{Exp2_results}
\begin{tabular}{|c|c|c|c|c|c|c|}
\hline
& $L_b$   & WI      & MWI    & Myopic & Rollout  & Rollout  \\
&  &  &  &  & ($H=3$) &  ($H=5$)   \\ \hline
 Value & $65.7$  & $64.3$  & $61.4$ & $63.5$ & $64.77$  & $63.76$ \\ \hline
\end{tabular}
\end{table}

\section{Conclusion}
\label{sec:conc}
In this paper, the problem of constrained restless multi-armed bandits is studied. These constraints are in the form of time varying availability of arms. The solution methods studied include Whittle's index policy, online rollout policy and myopic policy. 

For Whittle's index policy, index computation is done offline and the indices are used for online decision making. Whereas the implementation of online rollout policy is entirely online. 
Complexity analysis shows that rollout policy with a short look ahead is less complex than Whittle's index policy. However, there is a trade off between offline computation and  online decision time.
Numerical experiments show that when only one arm is played per slot ($M=1$), the online rollout policy is almost optimal, and is followed closely by the Whittle's index policy. This suggests that the rollout policy with a short look ahead can be used as an alternative to Whittle's index policy under computational scarcity. 
Further, as arms played per slot increases the performance of these policies seem to get closer to each other. This suggests that myopic policy might be `good enough' where larger number of arms are played.

A useful research direction would be to study variations on online rollout policy such as using different base policies.
Another future direction could be towards developing learning algorithms for scenarios where the systems parameters are unknown. 

%

\bibliographystyle{IEEEbib}
\bibliography{limfee}

\section*{Appendix}
\label{appendix}
\subsection{Proof of Lemma~\ref{lemma:valfuncprops}}
\label{proof:valfunc_convexity}
\begin{IEEEproof}
The proof uses the principle of mathematical induction.
\begin{enumerate}
\item First prove convexity of value functions in $\pi$ for fixed $w.$

Let  $V_1(\pi,1) := \max\{ \pi\eta_0 + (1-\pi)\eta_1, w \},$  and $V_1(\pi,0) :=  w.$ Observe that $V_1(\pi,1)$ is convex function in $\pi$ and $V_1(\pi,0)$ is constant function in $\pi;$ hence convex in $\pi.$ 

We next assume $V_n(\pi,1),$ and $V_n(\pi,0)$ are  convex in $\pi.$ The action value functions are given by 
{\footnotesize{
\begin{align*}
\mathcal{ L}^{1}V_{n}(\pi  ,1)= \eta(\pi) +   \beta
\rho(\pi)\left[ \theta^{1}_{1}(\pi)V_n(\Gamma_{1}(\pi),1)   +   (1-\theta^1_1(\pi))  \right. \\ \left. V_n(\Gamma_{1}(\pi),0) \right] + 
 \beta(1-\rho(\pi))\left[ \theta^1_{1}(\pi)V_n(\Gamma_{0}(\pi),1) + \right. \\ \left.  
  (1-\theta^1_{1}(\pi))V_n(\Gamma_{0}(\pi),0) \right],
\end{align*} 
\begin{align*}
\mathcal{L}^0V_{n}(\pi,1) = w + \beta
\left[ \theta^0_1(\pi) V_n(\gamma^0_1(\pi),1) +
  (1-\theta^0_1(\pi)) \right. \\ \left.
  V_n(\gamma^0_1(\pi),0) \right],
\end{align*}
\begin{align*}
\mathcal{L}^0V_{n}(\pi,0) = w + \beta\left[\theta^0_{0}(\pi) V_{n}(\gamma^0_0(\pi),1) +  (1-\theta^0_0(\pi))  \right. \\ \left.
   V_{n}(\gamma^0_0(\pi),0) \right].       
\end{align*}
}}
We now want to show that the above value functions are convex in $\pi.$ Convexity is not obvious from preceding equations, so we will rearrange the terms. Define
{\footnotesize{
\begin{align*}
b_0 := \left[ \pi p_{00} (1-\rho_0) + (1-\pi)p_{10}(1-\rho_1),   \pi (1-p_{00}) (1-\rho_0) 
\right. \\ \left. 
+ (1-\pi)(1-p_{10})(1-\rho_1)  \right],
\end{align*}
\begin{align*}
b_1 := \left[ \pi p_{00} \rho_0 + (1-\pi)p_{10}\rho_1,  \pi (1-p_{00})\rho_0 + 
\right. \\ \left.
 (1-\pi)(1-p_{10})\rho_1  \right],
\end{align*}
\begin{align*}
c_0 := \left[\pi p_{00} + (1-\pi) p_{10},\text{  } \pi (1-p_{00}) + (1-\pi) (1-p_{10}) \right].
\end{align*}
\begin{eqnarray*}
\widehat{b}_0 := \theta^1_1(\pi)b_0, &  \widetilde{b}_0 := (1-\theta^1_1(\pi))b_0	 \\
\widehat{b}_1 := \theta^1_1(\pi)b_1, & \widetilde{b}_1 := (1-\theta^1_1(\pi))b_1  \\ 
\widehat{c}_1 :=\theta^0_1(\pi)c_0, & \widetilde{c}_1 := (1-\theta^0_1(\pi))c_0, \\
\widehat{c}_0 :=\theta^0_0(\pi)c_0, & \widetilde{c}_0 := (1-\theta^0_0(\pi))c_0.
\end{eqnarray*}
}}
After rearranging terms, we can rewrite the action value functions as follows. 
{\footnotesize{
\begin{align*}
\mathcal{ L}^{1}V_{n}(\pi ,1) &= \eta(\pi) + \\ & \hspace{-0.5cm} \beta\left[
\lVert \widehat{b}_1 \rVert_1V_n\left(\frac{\widehat{b}_1}{\lVert \widehat{b}_1 \rVert_1},1\right)+ \lVert \widetilde{b}_1 \rVert_1V_n\left(\frac{\widetilde{b}_1}{\lVert \widetilde{b}_1 \rVert_1},0\right) \right]  
\\
& \hspace{-0.5cm} + \beta \left[ \lVert\widehat{b}_0\rVert_1 V_n\left(\frac{\widehat{b}_0}{\lVert \widehat{b}_0 \rVert_1},1\right) + 
   \lVert \widetilde{b}_0 \rVert_1 V_n\left(\frac{\widetilde{b}_0}{\lVert \widetilde{b}_0 \rVert_1},0\right) \right],
\end{align*}
\begin{align*}
\mathcal{L}^0V_{n}(\pi,1) =  w + \beta
\left[ \lVert \widehat{c}_1 \rVert_1 V_n\left(\frac{\widehat{c}_1}{\lVert \widehat{c}_1 \rVert_1},1\right) +
\right. \\ \left. 
  \lVert \widetilde{c}_1 \rVert_1V_n\left(\frac{\widetilde{c}_1}{\lVert \widetilde{c}_1 \rVert_1},0\right) \right],
\end{align*}
\begin{align*}
\mathcal{L}^0V_{n}(\pi,0) =  w + \beta
\left[ \lVert \widehat{c}_0 \rVert_1 V_n\left(\frac{\widehat{c}_0}{\lVert \widehat{c}_0 \rVert_1},1\right) + 
\right. \\ \left.  
 \lVert \widetilde{c}_0 \rVert_1V_n\left(\frac{\widetilde{c}_0}{\lVert \widetilde{c}_0 \rVert_1},0\right) \right].
\end{align*}
}}
From \cite[Lemma~$2$]{astrom1969}, given a convex function $g(x),$ the  function $\lVert x \rVert_1 g\left({x}/{\lVert x \rVert}_1\right)$ is also convex. 
Hence $\mathcal{ L}^{1}V_{n}(\pi ,1),$ $\mathcal{ L}^{0}V_{n}(\pi ,1)$ and $\mathcal{ L}^{0}V_{n}(\pi ,0)$ are convex functions in $\pi.$ As $V_{n+1}(\pi ,1)=\max\{\mathcal{ L}^{1}V_{n}(\pi ,1),\mathcal{ L}^{0}V_{n}(\pi ,1)\},$ it is also convex in  $\pi.$ Similarly, $V_{n+1}(\pi ,0)$ is convex  function in $\pi.$ 
As $n\rightarrow \infty, $  $V_{n}(\pi,1)$ and $V_{n}(\pi,0)$ converges uniformly to  $V(\pi,1)$ and $V(\pi,0),$ respectively. Thus  $V(\pi,1)$ and $V(\pi,0)$ are convex in $\pi.$ Also,  $\mathcal{ L}^{a}_{w}V(\pi ,y)$ is convex in $\pi$ for fixed $w.$

\item Analogously, the value functions are convex in $w$ for fixed $\pi.$ The proof technique is similar to that presented above. 
\end{enumerate}
\end{IEEEproof}
\subsection{Proof of Lemma~\ref{lemma:derivativewrt_pi_bound}}
\label{proof:derivativewrt_pi_bound}
\begin{IEEEproof} 
	The proof makes use of the principle of induction. We write down the steps for the case $p_{00}>p_{10}.$ For the other case $(p_{00}>p_{10}),$ similar steps will lead to the result. 
	
	In the first step, let $V_1(\pi,1) = \max \{\eta(\pi) , w \},$ where $\eta(\pi) = \pi(\eta_0 - \eta_1) + \eta_1$ and $\eta_0 < \eta_1 .$ Hence, the absolute value of slope of $V_1(\pi,1)$ w.r.t. $\pi$ is bounded by $\kappa c(\rho_1 - \rho_0).$  Also, observe that the absolute value of slope of $V_1(\pi,0)$ w.r.t. $\pi$ is $0,$ thus bounded by $\kappa c(\rho_1 - \rho_0).$	
	
	Next we assume that $|\frac{\partial V_n(\pi,1)}{\partial \pi}| \leq \kappa c(\rho_1 - \rho_0),$ and compute the  partial derivatives of $\mathcal{L}^1V_{n}(\pi,1),$ $\mathcal{L}^0V_{n}(\pi,1)$ and $V_{n+1}(\pi,1)$ w.r.t. $\pi.$
	{\small{
		\begin{dmath*}
		\mathcal{L}^1V_{n}(\pi,1) = \eta(\pi) +\beta \left[ \rho (\pi )
		\left({\theta_1^1}(\pi) V_n({\Gamma_{1,1}^1}(\pi ),1) 
		+  (1 - \theta_1^1(\pi))
		\right.  \left. 
		 {V}_n({\Gamma_{1,1}^1}(\pi ),0)\right) \right. \\ \left.
		+ (1 - \rho (\pi )\left( {\theta_1^1(\pi)} V_n({\Gamma_{1,0}^1}(\pi ),1) 
		\right. \right.  \left. \left. 
		+ (1 - {\theta_1^1}(\pi))
		{V}_n({\Gamma_{1,0}^1}(\pi ),0))\right)\right], \\
		\mathcal{L}^0V_{n}(\pi,1) = w + \beta[ \theta^0_1(\pi) V_n(\gamma_1^0(\pi),1) + (1-\theta^0_1(\pi)) V_n(\gamma^0_1(\pi),0)].
		\end{dmath*} 
	}}
	Taking partial derivative w.r.t. $\pi,$ we have
	{\small{
	\begin{dmath}
		\frac{\partial \mathcal{L}^1 V_{n}(\pi,1)}{\partial \pi} = (\eta_0 - \eta_1) + \beta (\rho_0-\rho_1)\left\lbrace \theta_1^1(\pi) V_n(\Gamma_{1}(\pi ),1) + 
		(1 - \theta_1^1(\pi)) {V}_n(\Gamma_{1}(\pi ),0)\right \rbrace 
		- \beta (\rho_0-\rho_1)\left\lbrace \theta_1^1(\pi) V_n(\Gamma_{0}(\pi ),1) + 
		(1 - \theta_1^1(\pi)) {V}_n(\Gamma_{0}(\pi ),0) \right\rbrace	 \\
		+ \beta \rho(\pi) \left\lbrace \theta_1^1(\pi) \frac{\partial V_n(\Gamma_{1}(\pi),1)}{\partial \Gamma_{1}(\pi)}  + 
		(1 - \theta_1^1(\pi)) \frac{\partial V_n(\Gamma_{1}(\pi),0)}{\partial \Gamma_{1}(\pi)} \right\rbrace \frac{\partial\Gamma_{1}(\pi)}{\partial \pi}
		+ \beta (1-\rho(\pi)) \left\lbrace \theta_1^1(\pi) \frac{\partial V_n(\Gamma_{0}(\pi ),1)}{\partial \Gamma_{0}(\pi)}  + 
		(1 - \theta_1^1(\pi)) \frac{\partial V_n(\Gamma_{0}(\pi ),0)}{\partial \Gamma_{0}(\pi)} \right\rbrace \frac{\partial\Gamma_{0}(\pi)}{\partial \pi}  
		\\ +  \beta\rho(\pi)\left\lbrace V_n(\Gamma_{1}(\pi ),1) - {V}_n(\Gamma_{1}(\pi ),0) \right\rbrace\frac{\partial\theta^1_1(\pi)}{\partial\pi} 
		\\ + \beta(1-\rho(\pi))\left\lbrace V_n(\Gamma_{0}(\pi ),1) - {V}_n(\Gamma_{0}(\pi ),0) \right\rbrace\frac{\partial\theta^1_1(\pi)}{\partial\pi}. 
	\label{derivative1}
	\end{dmath} } }
	Assuming $\theta^a_y(\pi)$ is independent of $\pi,$ i.e., $\theta^a_y(\pi)= \theta^a_y,$ 	and using {\footnotesize{\[\frac{\partial\Gamma_{1}(\pi)}{\partial \pi} = \frac{\rho_0\rho_1(p_{00}-p_{10})}{(\rho(\pi))^2},\text{ }\frac{\partial\Gamma_{0}(\pi)}{\partial \pi} = \frac{(1-\rho_0)(1-\rho_1)(p_{00}-p_{10})}{(1-\rho(\pi))^2}\]}} 
	along with the fact $\rho(\pi)\in [\rho_0,\rho_1],$   $1-\rho(\pi)\in [1-\rho_1,1-\rho_0],$  we have  
	{\small{
	\begin{dmath}
		\frac{\partial \mathcal{L}^1 V_{n}(\pi,1)}{\partial \pi} \leq {(\rho_1-\rho_0)\bigg( -b + \beta\theta^1_1 \left[V_n(\Gamma_0(\pi),1) - V_n(\Gamma_1(\pi),1) \right] }+  \beta(1-\theta^1_1)\left[V_n(\Gamma_0(\pi),0) - V_n(\Gamma_1(\pi),0) \right] \bigg) + 
		\beta (p_{00}-p_{10})\left[ \left\lbrace \theta_1^1 \frac{\partial V_n(\Gamma_{1}(\pi),1)}{\partial \Gamma_{1}(\pi)}  + (1 - \theta_1^1) \frac{\partial V_n(\Gamma_{1}(\pi),0)}{\partial \Gamma_{1}(\pi)} \right\rbrace \rho_1
		+  \left\lbrace \theta_1^1 \frac{\partial V_n(\Gamma_{0}(\pi ),1)}{\partial \Gamma_{0}(\pi)}  + 
		(1 - \theta_1^1) \frac{\partial V_n(\Gamma_{0}(\pi ),0)}{\partial \Gamma_{0}(\pi)} \right\rbrace (1-\rho_0)\right] .
	\label{derivative1_upperbound}
	\end{dmath} 
	}}
By using the fact $\rho(\pi)\leq \rho_1$ and $1-\rho(\pi)\leq 1-\rho_0,$ we have 
{\small{
	\begin{dmath}
		\frac{\partial \mathcal{L}^1 V_{n}(\pi,1)}{\partial \pi} \geq {(\rho_1-\rho_0)\bigg( -b + \beta\theta^1_1 \left[V_n(\Gamma_0(\pi),1) - V_n(\Gamma_1(\pi),1) \right] }+  \beta(1-\theta^1_1)\left[V_n(\Gamma_0(\pi),0) - V_n(\Gamma_1(\pi),0) \right] \bigg) + 
		\beta(p_{00}-p_{10})\left[ \left\lbrace \theta_1^1 \frac{\partial V_n(\Gamma_{1}(\pi),1)}{\partial \Gamma_{1}(\pi)}  + (1 - \theta_1^1) \frac{\partial V_n(\Gamma_{1}(\pi),0)}{\partial \Gamma_{1}(\pi)} \right\rbrace \rho_0
		+  \left\lbrace \theta_1^1 \frac{\partial V_n(\Gamma_{0}(\pi ),1)}{\partial \Gamma_{0}(\pi)}  + 
		(1 - \theta_1^1) \frac{\partial V_n(\Gamma_{0}(\pi ),0)}{\partial \Gamma_{0}(\pi)} \right\rbrace (1-\rho_1)\right].
		\label{derivative1_lowerbound}
	\end{dmath} 
	}}	
	 For $p_{00} > p_{10}$ we have   $p_{10} \leq \Gamma_{1}(\pi), \Gamma_{0}(\pi) \leq p_{00}.$ From our assumption on $V_n$ we  obtain  
{\small{\begin{dmath}
	\lvert V_n(\Gamma_1(\pi),1)- V_n(\Gamma_0(\pi)) \rvert \leq \kappa c (\rho_1-\rho_0) \lvert \Gamma_1(\pi)- \Gamma_0(\pi) \rvert \leq \kappa c (\rho_1-\rho_0)\lvert p_{00}-p_{10} \rvert . \label{lips-difference-bound}
	\end{dmath} }}
	Substituting in \eqref{derivative1_upperbound}, we get 
	{\small{
    \begin{dmath*}
    	\frac{\partial \mathcal{L}^1 V_{n}(\pi,1)}{\partial \pi} \leq (\rho_1-\rho_0)\left\lbrace -b + \beta\kappa  c(\rho_1-\rho_0)(p_{00}-p_{10}) + \beta\kappa c\rho_1(p_{00}-p_{10}) + \beta\kappa c(1-\rho_0)(p_{00}-p_{10}) \right\rbrace\\
    	%
    	\leq (\rho_1 - \rho_0)\{-b + 3\beta\kappa c(p_{00}-p_{10}) \}
    	\leq \kappa c (\rho_1 - \rho_0)\{ -b + \beta(b+3c)(p_{00}-p_{10}) \}. 
	\end{dmath*} }}
Using $\frac{\partial V_n(\pi,1)}{\partial \pi}\geq -\kappa c (\rho_1-\rho_0)$ and $ V_n(\Gamma_1(\pi),1)- V_n(\Gamma_0(\pi)) \geq -\kappa c (\rho_1-\rho_0)\lvert p_{00}-p_{10} \rvert $ in \eqref{derivative1_lowerbound}, we have
{\small{
\begin{dmath*}
\frac{\partial \mathcal{L}^1 V_{n}(\pi,1)}{\partial \pi} \geq (\rho_1-\rho_0)\left\lbrace -b - \beta\kappa  c(\rho_1-\rho_0)(p_{00}-p_{10}) - \beta\kappa c\rho_0(p_{00}-p_{10}) - \beta\kappa c(1-\rho_1)(p_{00}-p_{10}) \right\rbrace\\
%
%
\geq -\kappa  (\rho_1 - \rho_0)\{b + \beta (p_{00}-p_{10})(c-b)\} \geq -\kappa c  (\rho_1 - \rho_0). 
\end{dmath*} }}
Under the condition $0<p_{00}-p_{10}<\frac{b+1}{b+3c},$ the expression $\lvert -b + \beta(b+3c)(p_{00}-p_{10})\rvert<1$ and the value of $\left|\frac{\mathcal{L}^1V_{n}}{\partial\pi}\right|$ is bounded by  $\kappa c(\rho_1 - \rho_0).$ By induction $\left|\frac{\partial\mathcal{L}^1V}{\partial\pi}\right|$ is also bounded by the same. 
	Similarly, for the case $p_{00}<p_{10},$ we have
	{\small{
			\begin{dmath*}
				\frac{\partial \mathcal{L}^1 V_{n}(\pi,1)}{\partial \pi} \leq (\rho_1-\rho_0)\left\lbrace -b + \beta\kappa c (\rho_1-\rho_0)(p_{10}-p_{00})\right\rbrace \\
				+ \beta \left\lbrace \theta_1^1(\pi) \frac{\partial V_n(\Gamma_{1}(\pi),1)}{\partial \Gamma_{1}(\pi)}  + 
				(1 - \theta_1^1(\pi)) \frac{\partial V_n(\Gamma_{1}(\pi),0)}{\partial \Gamma_{1}(\pi)} \right\rbrace \frac{\rho_0\rho_1(p_{00}-p_{10})}{\rho(\pi)}
				+ \beta \left\lbrace \theta_1^1(\pi) \frac{\partial V_n(\Gamma_{0}(\pi ),1)}{\partial \Gamma_{0}(\pi)}  + 
				(1 - \theta_1^1(\pi)) \frac{\partial V_n(\Gamma_{0}(\pi ),0)}{\partial \Gamma_{0}(\pi)} \right\rbrace \frac{(1-\rho_0)(1-\rho_1)(p_{00}-p_{10})}{(1-\rho(\pi))}
				%
				%
				%
				%
				\leq \kappa c (\rho_1-\rho_0) \{ -b + \beta (b+c)(p_{10}-p_{00})\}.
			\end{dmath*}
	}}
	Under the condition $0<p_{10}-p_{00}<\frac{b+1}{b+c},$ the expression $\lvert -b + \beta(b+c)(p_{10}-p_{00}\rvert <1$ and the value of $\left| \frac{\partial\mathcal{L}^1V_{n}}{\partial \pi}\right|$ is bounded by  $\kappa c(\rho_1 - \rho_0).$ By induction $\left|\frac{\partial\mathcal{L}^1V}{\partial\pi}\right|$ is also bounded by the same. 
	
	We now want to bound $\left|\frac{\partial\mathcal{L}^0V_{n}(\pi,1)}{\partial\pi}\right|.$
	Again using induction, assume that
	 $\left|\frac{\partial V_{n}(\pi,1)}{\partial\pi}\right|=0.$ Assume $\left|\frac{\partial\mathcal{L}^0V_{n}(\pi,1)}{\partial\pi}\right|\leq\kappa c (\rho_1-\rho_0).$ 
	Taking partial derivative of $\mathcal{L}^0V_{n}(\pi,1)$ w.r.t. $\pi$ we get 
	{\small{ 
			\begin{dmath*}
				\frac{\partial\mathcal{L}^0V_{n}(\pi,1)}{\partial\pi} =  \beta\left[\theta^0_1(\pi)\frac{\partial V_n(\gamma^0_1(\pi),1)}{\partial\pi} + (1-\theta^0_1(\pi))\frac{\partial V_n(\gamma^0_1(\pi),0)}{\partial\pi} \right]\frac{\partial\gamma^0_1(\pi)}{\partial\pi} 
				+ \beta \left[ V_n(\gamma^0_1(\pi),1) - V_n(\gamma^0_1(\pi),0) \right] \frac{\partial \theta^0_1(\pi)}{\partial\pi}.
			\end{dmath*}
		}}
	
		For $\theta^0_1(\pi) = \theta^0_1,$ after simplification we get    
		 $
		  {\left| \frac{\partial\mathcal{L}^0V_{n}(\pi,1)}{\partial\pi} \right| \leq \beta\kappa c(\rho_1-\rho_0)\left|p_{00}-p_{10}\right|.}
		  $ 

By induction $\left| \frac{\partial\mathcal{L}^0 V_n}{\partial \pi}\right|$ is bounded by  $\kappa c(\rho_1 - \rho_0).$ As $V_{n+1}(\pi,1) = \max \{ {\mathcal{L}^1V_{n}}(\pi,1),{\mathcal{L}^0V_{n}}(\pi,1)\},$ the derivative of $V_{n+1}(\pi,1)$ is also bounded by $\kappa c (\rho_1-\rho_0).$ 	
As $n \rightarrow \infty,$ we get $V_n(\pi,1) \rightarrow V(\pi,1)$ and  $V_n(\pi,0) \rightarrow V(\pi,0)$ uniformly.  Hence we have desired bound on partial derivative of value functions. 
\end{IEEEproof}
\subsection{Proof of Lemma~\ref{lemma:advantage}}
\label{proof:advantage}
{
\begin{IEEEproof}
	To show that $D(\pi)$ is decreasing, it is enough to show that $\frac{\partial D(\pi)}{\partial \pi} < 0.$ 
	{\footnotesize{
			\begin{dmath*}
				\frac{\partial D(\pi)}{\partial \pi} = \frac{\partial\mathcal{L}^1V(\pi,1)}{\partial\pi} - \frac{\partial\mathcal{L}^0V(\pi,1)}{\partial\pi}.
	\end{dmath*} }}
	Using Lemma~\ref{lemma:derivativewrt_pi_bound}, for the case $p_{00}>p_{10},$
	{\small{
	\begin{dmath*}
			\frac{\partial D(\pi)}{\partial \pi} \leq \kappa c(\rho_1-\rho_0)\left\lbrace -b + \beta(b+3c)(p_{00}-p_{10})\right\rbrace + \beta\kappa c(\rho_1-\rho_0)(p_{00}-p_{10})\\
				\leq \kappa c(\rho_1-\rho_0)\left\lbrace -b + \beta(b+3c+1)(p_{00}-p_{10})\right\rbrace .
	\end{dmath*} }}
	Under the condition $0<p_{00}-p_{10}<\frac{b}{b+3c+1},$ the coefficient $-b + \beta(b+3c+1)(p_{00}-p_{10})$ is negative; hence, the derivative $\frac{\partial D(\pi)}{\partial \pi}$ is negative.\\
	Similarly, for the case $p_{00}<p_{10},$ 
	{\small{
	\begin{dmath*}
		\frac{\partial D(\pi)}{\partial \pi} \leq \kappa c(\rho_1-\rho_0)\left\lbrace -b + \beta(b+c)(p_{10}-p_{00})\right\rbrace + \beta\kappa c(\rho_1-\rho_0)(p_{10}-p_{00})\\
		\leq \kappa c(\rho_1-\rho_0)\left\lbrace -b + \beta(b+c+1)(p_{10}-p_{00})\right\rbrace .
	\end{dmath*} }}
	Under the condition $0<p_{10}-p_{00}<\frac{b}{b+c+1},$ the coefficient $-b + \beta(b+c+1)(p_{00}-p_{10})$ is negative; hence, the derivative $\frac{\partial D(\pi)}{\partial \pi}$ is negative.\\
\end{IEEEproof}
}
%
\subsection{Proof of Lemma \ref{lemma:Lagrangian_decouple} }
\label{proof:Lagrangian_decouple}
\begin{IEEEproof}
{
Denote the expressions on the right hand side of \eqref{eqn:crmab-Lag_relaxed} and  \eqref{eqn:lemma-Lagrangian_decouple} as $\mathcal{E}_1$ and $\mathcal{E}_2$ respectively. We need to show that substituting $\mathcal{E}_2$ in \eqref{eqn:crmab-Lag_relaxed} gives \eqref{eqn:lemma-Lagrangian_decouple}, i.e. $\mathcal{E}_1(\mathcal{E}_2) = \mathcal{E}_2.$ That means, it suffices to show that the following expression $\mathcal{E}_1(\mathcal{E}_2)- \mathcal{E}_2$ equals $0.$}
{\small{
\begin{dmath*}
\max\limits_{\bm{a}\in \mathcal{A}_{\bm{y}}} \bigg\lbrace\sum\limits_{n=1}^{N} [r_n(\pi_n,y_n,a_n)-\lambda a_n] + \lambda M + \beta\sum\limits_{\bm{o}\in S_{\bm{o}}}\sum\limits_{\bm{y'}\in S_{\bm{y}}} \prob{\bm{o}|\bm{\pi},\bm{y},\bm{a}}\prob{\bm{y'}|\bm{y},\bm{a}}\bigg[\frac{M\lambda}{1-\beta} + \sum\limits_{n=1}^{N}J^{\lambda}(\bm{\Gamma}^{o_n}({\pi}_n),{y'_n}) \bigg] \bigg\rbrace - \frac{M\lambda}{1-\beta} - \sum\limits_{n=1}^{N}J^{\lambda}({\pi}_n,{y_n}), \\
\text{using {\footnotesize{$\sum\limits_{\bm{o}\in S_{\bm{o}}}\sum\limits_{\bm{y'}\in S_{\bm{y}}} \prob{\bm{o}|\bm{\pi},\bm{y},\bm{a}}\prob{\bm{y'}|\bm{y},\bm{a}}=1,$}}  }\\
\text{\hfill and rearranging the terms,}\\
= - \sum\limits_{n=1}^{N}J^{\lambda}({\pi}_n,{y_n}) + \max\limits_{\mathcal{A}_{\bm{y}}} \bigg\lbrace\sum\limits_{n=1}^{N} [r_n(\pi_n,y_n,a_n)-\lambda a_n] + \beta\sum\limits_{\bm{o}\in S_{\bm{o}}}\sum\limits_{\bm{y'}\in S_{\bm{y}}}\sum\limits_{n=1}^{N} \prob{\bm{o}|\bm{\pi},\bm{y},\bm{a}}\prob{\bm{y'}|\bm{y},\bm{a}}  J^{\lambda}(\bm{\Gamma}^{o_n}({\pi}_n),{y'_n})  \bigg\rbrace , \\
\text{reordering the summations and suitably expanding, }\\
= - \sum\limits_{n=1}^{N}J^{\lambda}({\pi}_n,{y_n}) + 	\max\limits_{\bm{a}\in \mathcal{A}_{\bm{y}}}  \bigg\lbrace \sum\limits_{n=1}^{N} [r_n(\pi_n,y_n,a_n)-\lambda a_n] \\ + \beta\sum\limits_{n=1}^{N}\sum\limits_{{o_n}\in S_{o_n}}\sum\limits_{{y'_n}\in S_{y'_n}}\sum\limits_{\bm{o}_{-n}\in S_{\bm{o}_{-n}}}\sum\limits_{\bm{y'}_{-n}\in S_{\bm{y}_{-n}}}  \left[ \prob{\bm{o}|\bm{\pi},\bm{y},\bm{a}} \times \\ \hspace{1.2cm}\prob{\bm{y'}|\bm{y},\bm{a}}  J^{\lambda}(\bm{\Gamma}^{o_n}({\pi}_n),{y'_n})	\right]	\bigg\rbrace ,
\end{dmath*} where, $\bm{o}_{-n}$ is the observation vector $\bm{o}$ omitting the $n^{th}$ element. So is the case with $\bm{y'}_{-n}$ and so on.
\begin{dmath*}
= - \sum\limits_{n=1}^{N}J^{\lambda}({\pi}_n,{y_n}) + \max\limits_{\bm{a}\in \mathcal{A}_{\bm{y}}}  \bigg\lbrace \sum\limits_{n=1}^{N} [r_n(\pi_n,y_n,a_n)-\lambda a_n]  +  \beta\sum\limits_{n=1}^{N}\sum\limits_{{o_n}\in S_{o_n}}\sum\limits_{{y'_n}\in S_{y'_n}} \left[ \prob{{o}_n|{\pi}_n,y_n,a_n} \times \\ \hspace{1.2cm}\prob{y'_n|y_n,a_n} J^{\lambda}(\bm{\Gamma}^{o_n}({\pi}_n),{y'_n}) \right]\bigg\rbrace \\
= { \sum\limits_{n=1}^{N}\bigg( -J^{\lambda}({\pi}_n,{y_n}) + \max\limits_{a_n\in \mathcal{A}_{y_n}}  \bigg\lbrace[r_n(\pi_n,y_n,a_n)-\lambda a_n] }\\ + \beta\sum\limits_{{o_n}\in S_{o_n}}\sum\limits_{{y'_n}\in S_{y'_n}} \left[ \prob{{o}_n|{\pi}_n,y_n,a_n} \times \\ \prob{y'_n|y_n,a_n}  J^{\lambda}(\bm{\Gamma}^{o_n}({\pi}_n),{y'_n})\right]\bigg\rbrace\bigg)\\=0.
\end{dmath*} }} 
\end{IEEEproof}

\end{document}